\def\rmn{\everymath={\fam0 }\fam0 } \def\hii{H~II}
\def\UBVI{\hbox{$U\!BV\!I$}} \def\BV{\hbox{$B\!-\!V$}}
\def\UB{\hbox{$U\!-\!B$}} \def\UV{\hbox{$U\!-\!V$}}
\def\VI{\hbox{$V\!-\!I$}} \def\ebv{E(\hbox{$B\!-\!V$})}
\def\V{$V$} \def\B{$B$} \def\U{$U$} \def\I{$I$}
  \def\etal{{\it
et al.\/\ }}  \def\eg{{\it e.g.\/}}

\documentstyle[11pt,aaspp4]{article}

\slugcomment{\sf Scheduled to appear in the September 1996 issue of The Astronomical Journal}

\lefthead{Bresolin, Kennicutt and Stetson} 
\righthead{OB associations and clusters in M101}

\begin{document}

\title{AN HST STUDY OF OB ASSOCIATIONS AND\\STAR CLUSTERS IN
M101~\footnote{ Based on observations with the NASA/ESA {\it
Hubble Space Telescope}, obtained at the Space Telescope
Science Institute, which is operated by AURA, Inc., under
NASA contract NAS 5-26555}}

\author{Fabio Bresolin, Robert C. Kennicutt, Jr.}
\affil{Steward Observatory, University of Arizona, Tucson, AZ 85721.
E-mail: fabio@as.arizona.edu, robk@as.arizona.edu}
\and
\author{Peter B. Stetson}
\affil{Dominion Astrophysical Observatory, 5071 West Saanich Road,
Victoria, BC, V8X 4M6, Canada. E-mail: stetson@dao.nrc.ca}

\begin{abstract} 
The massive stellar content, the OB associations and the star clusters 
in an HST field in M101 = NGC~5457 are investigated.  A clustering 
algorithm yields 79 putative associations.  Their size distribution is 
similar to that found in the Magellanic Clouds, M31 and M33, with an 
average size around 90 pc.  The \V\ luminosity function for the stars 
contained within the associations has a slope $d\,$log$N$/$d\,V = 0.60 
\pm 0.05$, while an average reddening \ebv\ = 0.21 mag is measured.  The 
stellar content is further discussed by means of color-magnitude and 
color-color diagrams.  Ages are estimated using theoretical 
isochrones, and range between 3 and 14 Myr ($\pm 2$ Myr).  We find a 
suggestion that the upper mass limit of the IMF for stars in OB 
associations in M101 may be quite high, contrary to some theoretical 
expectations that the mass limit should be lower in a high metallicity 
environment.  Forty-one star cluster candidates and two H~II region 
core clusters are identified in the M101 field, and their integrated 
photometric properties are compared with the cluster system of the LMC 
and M33.  Most of the M101 clusters probably belong to the class of 
young, populous star clusters such as are found in the LMC. Red 
clusters are rare in this field.  In the Appendix the objective 
finding algorithm is applied to the brightest stars in the Large 
Magellanic Cloud.

\end{abstract}


\section{INTRODUCTION}

OB associations provide valuable information on the physical processes
that govern star formation in galaxies.  Unlike star clusters, which
are gravitationally bound systems consisting of presumably coeval
stars, OB associations are loose, short-lived entities, and as such
they are natural tracers of recent or current star formation
(\cite{blaauw91}). A better understanding of massive stars has
resulted from the investigation of these objects in the Galaxy and in
the Magellanic Clouds (\cite{garmany94}).  The study of the spatial
distribution of OB associations can provide an insight into the mode
of star formation (stochastic vs.  density wave triggering), while the
size distribution of the associations is directly related to the
question of the importance of the associations and star complexes as
fundamental building blocks of the structure of spiral galaxies
(\cite{efremov94}). On the other hand star clusters hold clues about
galactic evolution on longer timescales.

The main advantage of studying OB associations in galaxies other than
the Milky Way lies in the possibility of minimizing the difficulties
that arise from uncertain distances and high levels of obscuration in
the plane of the Galaxy.  However we must deal with the long-standing
problem of identifying stellar associations in a consistent way for
different galaxies. \cite{hodge86} pointed out how the measured
association properties depend on several selection effects, among
these the quality of the observational material and the identification
criteria adopted.  To overcome this problem {\it automated}
identification techniques have been recently applied to nearby,
resolved galaxies.  In this work, which introduces a project aimed at
defining some of the properties of OB associations and star clusters
in different galactic environments (metallicity, star forming
activity, etc.), we present the results obtained for M101 with a
similar objective technique, and describe the methodology adopted for
investigating associations and clusters in external galaxies.  Our goal is to
analyze galaxies of different Hubble types and distances (up to 15 Mpc
with HST), in order to compare their OB associations, their star
clusters, and their massive stellar content.  This can provide clues
on the possible variations of stellar populations and massive star
formation among galaxies.

We present the observational material and the data reduction in \S~2.  
Results on the massive stars are discussed in \S~3.  The identification 
technique and the size distribution of the OB associations in M101 are 
discussed in \S~4 and \S~5, respectively.  In \S~6 we analyze the 
properties of the stellar content of the associations.  In \S~7 we draw 
some tentative conclusions on the stellar IMF in the field studied.  The 
results on star clusters are presented in \S~8.  In the Appendix we apply 
the objective algorithm to the LMC.  A distance to M101 of 7.4 Mpc 
(\cite{kelson96}) is adopted.

\section{OBSERVATIONS AND DATA REDUCTION} 
The HST Distance Scale Key Project is observing 18 galaxies to detect
Cepheid variables (\cite{kfm}). As a by-product of this effort, 
a large number of WFPC2 images of
spiral galaxies is produced, suitable for the study of their stellar
populations.  Figure~\ref{mosaic} shows the inner field of M101 = NGC~5457
studied in this work (see also \cite{stetson96}).  The apex of the four
chips lies at $\rmn \alpha = 14^h03^m24\fs0$ and
$\delta$~=~$54\arcdeg21\arcmin38\arcsec$ (2000.0), $1\farcm 8$ NE of
the center of M101, as shown in Fig.~\ref{halphaimage}.  The filters
used for the observations were F336W (two 1200s exposures), F439W (two
exposures, 1000s and 1200s respectively), F555W (12 1200s exposures)
and F814W (one 1000s and four 1200s exposures). 
The stellar photometry was carried out with the multiple-frame
profile-fitting package ALLFRAME (\cite{stetson94}). 
In order to
create a master list of stellar objects, a median image of each chip,
free of cosmic rays, was generated from all the available single-epoch
images.  Once the stellar list was made using the PSF-fitting program
ALLSTAR, we performed the photometry for each of the single-epoch
images with ALLFRAME.  The average magnitudes were calibrated and
transformed to the \UBVI\ system with the equations and zero points of
\cite{holtzman95}.  For the following analysis only stars that were
measured in all four filters were retained, giving a total of
16812 stars.

Table~\ref{errors.table} lists the typical photometric errors at
various magnitudes.  The exposure times for the \B\ and \U\ frames
were not long enough to accurately measure stars fainter than
\V$\sim$24.  This imposes some limitations on the analysis of the
fainter stars, but not on the brighter ones which are of interest
here.  The incompleteness in the \V\ photometry was estimated by
reducing copies of the \V\ frames of chip 4 (the most crowded) to
which 1,000 artificial stars were added.  Virtually all stars down to
\V$ = 25$ were recovered, 93\% at
\V$=25.5$, 80\% at \V$=26$. However, in our study of the
luminosity function (\S~\ref{LF}), incompleteness is seen to set in
at \V$\sim$23.5 and appears to be more severe than stated above at
fainter magnitudes.  Since only stars recovered in all four filters
were included, this can be ascribed to two causes: shorter exposure
times in \B\ and \U, and the fact that the artificial stars were added
at completely random places within the field, whereas the majority of
real stars are -- by definition -- located in regions of
higher-than-average stellar density, and are therefore more subject to
crowding and confusion.  Approximate completeness limits, without
restriction to stars detected in all four filters, are
\V$\sim$23.5 in \U\ and \B\, and \V$\sim$25 in \V\ and \I.

Ground-based images at H$\alpha$ and H$\beta$ of a region partly
overlapping with the HST field were secured in March 1994 with the
Steward Observatory 90-inch telescope on Kitt Peak.
These data are used in \S~\ref{reddening} to independently estimate
the reddening in the corresponding OB associations.

\section{MASSIVE STARS}

Figure~\ref{tracks.com} shows the color-magnitude diagram (CMD) for
$\sim$16,000 stars recovered in the field studied.  The evolutionary
tracks, taken from \cite{schaller92}, were converted to the
observational plane using the equations of \cite{massey95}. They were
reddened by E($B\!-\!V$)~=~0.21, corresponding to the mean extinction measured
for the stars.  This diagram shows that we are
detecting stars typically above 10 M$_\odot$. The width of the blue
plume is $\sim$0.5 magnitudes for \V$<$24.5, and increases to one
magnitude at the faint end. This is much larger than what is observed
in OB associations in the LMC (Massey \etal 1989, \cite{hunter95b}). 
The stellar models of \cite{schaller92} predict a wide main sequence
for the H-R diagram of massive stars. However, given the photometric
uncertainties (Table~\ref{errors.table}), the observed scatter is
largely explained by photometric errors alone, and perhaps partially
by an age spread among non-coeval populations of stars and by
differential reddening. Similar conclusions were drawn by
\cite{hunter95} for HST data of I Zw 18. Effects of blending are
surely present, as many stars are located in crowded regions, so that
several of the brightest objects could in fact be groups rather than
single stars.  To reduce this effect for the brightest objects, we
have removed from the CMD all stars brighter than \V~=~22 that do not
appear to be isolated stars. The result is shown in the inset of
Fig.~\ref{tracks.com}, where the photometric errors for each star are
indicated. 

\section{IDENTIFICATION OF OB ASSOCIATIONS\label{identification}}

Recently the question of comparing the clustering properties of massive
stars in different galaxies has been attacked by adopting automated
algorithms, to remove much of the subjectivity which was intrinsic to
previous methods. \cite{wilson91} introduced a friends-of-friends
algorithm to define groups of bright blue stars in M33. In this method
all blue stars lying within a predetermined search radius from another
star were included in the same stellar group. The latter was defined
as an OB association if it contained at least 10 blue stars.  The
search radius was determined by the mean surface density of blue
stars. A somewhat different method, introduced by \cite{battinelli91},
determines the search radius that maximizes the number of stellar
groups containing a minimum of three stars. The same method has been
used by Magnier \etal (1993) to identify OB associations in M31.

In this study of M101 we adopted Battinelli's (1991) approach, because
of its high level of objectivity and in order to compare association
properties of galaxies already studied. The criteria \V $<$ 24.5
(M$_V$ $<$ $-$4.8) and (\BV) $<$ 0.4 (before correcting for reddening)
were used to select the blue stars. The search algorithm was applied
separately to the four WFPC2 chips because of the gradient in stellar
density across the HST field. This gave search radii of 2\farcs4 (40
pc), 3\farcs5 (58 pc), 3\farcs9 (64 pc) and 2\farcs2 (36 pc) in chips
1, 2, 3 and 4, respectively. These values are very close (to within
$\pm$ 0.2 arcsec) to what one obtains using Wilson's (1991) definition
of search radius based on the stellar surface density.
We must stress that the `associations' defined by the algorithm lie
within a two-dimensional projection through a patch of galactic disk
or a spiral-arm fragment, and do not {\it necessarily} lie physically
close together in three dimensions, share common space motions, or
originate in a single coherent star-forming event; thus we cannot be
certain that any given one of them is a true OB association as we
would define it in our Milky Way Galaxy.  Nevertheless, even though we
cannot be sure that these asterisms are genuine, physical OB
associations, we can at least be confident that they represent
objectively defined, localized regions of enhanced surface density in
the distribution of young, massive stars.

We can perform a statistical test to ascertain how many of the groups
found are likely to be chance coincidences. We applied the algorithm
to a set of random distributions of stars, each containing the same
number of blue stars as the actual data.  The search radius was set equal
to the value used for the actual data.  The simulations showed that
the contamination of false associations was less than 10$\%$
when the threshold population was set to seven stars or more. We
adopted this criterion, which resulted in a total sample of 79
associations, as shown in Figure~\ref{clust.com}. Fifty contain at
least 10 blue stars.  The average number of blue stars in the
associations is 15 (19 for $\rmn N_{blue} \ge 10$). It is likely that
the actual number of member stars is larger than these figures, given the
possibility of measuring very compact groups of a few
stars as a single star. The simulations described do not take into account the
fact that young stars are not distributed exactly at random due to the
existence of spiral arms and dust lanes.

Our sample is biased against associations that contain very few stars.
A major concern is therefore the completeness in the rate of detection
of associations. To estimate the incompleteness we made a comparison
with the system of associations in the LMC.  The area covered by our
HST field is approximately equal to the area surveyed by Lucke and
Hodge (1970) in the LMC, where 122 associations were catalogued.
Correcting for inclination effects and scaling to the number of LMC
associations, we should have detected $\sim$100 associations in the
M101 field. We might therefore be incomplete by 20\%, on the
assumption that the two galaxies are directly comparable.  Appendix~A
shows however that when the objective algorithm is applied to the LMC,
the same number of associations is detected as were found in the works
of Lucke and Hodge (1970).  A possible cause for incompleteness arises
from the difficulty in establishing very small groups of blue stars as
real associations.  Our search method regards most of the smallest
groups as statistically insignificant, but many of them could be very
compact associations, for which we are not sensitive enough. This
means that we are not detecting ``Orion-like'' objects in
M101. Instead the smallest objects in our sample correspond to
Galactic star forming regions with size between that of M8 and the
Rosette Nebula (NGC~2244), both in terms of diameter and H$\alpha$
luminosity (\cite{kennicutt84}).  At the high-end we find Galactic
counterparts in the Carina and W49 regions, while the most luminous
association (\# 72) has an H$\alpha$ luminosity comparable to 30
Doradus in the LMC.  On the other hand we think that our search method
is conservative, in the sense of disregarding associations of dubious
reality, which will be very important when studying galaxies at even
larger distances.  We are therefore at a necessary trade-off between
completeness and reproducibility.

Since we are interested in comparing results for different galaxies in
the future, we ran a simple experiment to test how resolution might
affect the identification of the associations. New versions of chip 4
images were created, compressed by a factor of 1.5 and 2, to simulate
a corresponding increase in distance (the exposure times were assumed
increased by a factor of 2.25 and 4.0 respectively).  The search radii
were reduced by exactly the same factors with respect to the original
images. The average size of the clumps did not change significantly in
the first case, and increased by 10\% in the case corresponding to a
doubling of the distance. In both cases, most of the original
morphology and position of the associations were recovered, while the
number of blue stars was reduced by 9\% and 19\%. These results
suggest that we can confidently compare properties of associations in
galaxies which differ by at least a factor of two in distance.

\section{SIZE DISTRIBUTION OF THE OB ASSOCIATIONS\label{size}}

The diameters of the associations found in M101 are given in
Table~\ref{obdata.table}, while Figure~\ref{size.com} compares the
size distribution of the M101 associations with other nearby galaxies
studied with similar methods.  These are the LMC (Appendix), M33
(Wilson 1991 and Regan and Wilson 1993), M31 (\cite{haiman94}), the
SMC (Battinelli 1991), and NGC~6822 (\cite{wilson92}).  Somewhat
different criteria have been used to select the blue stars in these
works; our criterion is equivalent to that applied to M33. We note
that the size distributions are dependent on the search radius chosen.
All galaxies show a similar distribution, and the Kolgomorov--Smirnov
test indicates that the data are indeed consistent with a single
distribution function. The presence of a peak might at first be
attributed to a selection effect, due to the fact that smaller
associations are more difficult to detect. However, an absence of
associations smaller than 20-30 pc is observed in even well resolved
galaxies such as the Magellanic Clouds (Hodge and Lucke 1970, Hodge
1985), indicating that the observed distributions are indeed
peaked. Furthermore in the Galaxy OB associations have a mean size
around 140 pc (Garmany and Stencel 1992), much larger than the typical
diameters of open clusters (10--35 pc, \cite{janes88}).  The peak in
the distributions lies around 40--80 pc for M101, LMC, SMC, M33 and
NGC~6822, but $\sim$110 pc for M31. This hint of a possible
Hubble-type dependence deserves further investigation. However, it
should be noted that the size distribution that Efremov \etal (1987)
obtained for the M31 associations shows a behavior more closely
resembling that of the remaining galaxies.

We believe the size distribution to be more meaningful than the mean
diameter, which is difficult to define and is subject to numerous
observational biases (see discussion in Magnier \etal 1993).  We have
however compiled in Table~\ref{compare.table} the mean for the
galaxies studied, which show similar values of $\sim$90 pc.  We
conclude that the associations in M101, LMC, SMC, M33, NGC~6822 and perhaps M31
have similar size distributions and average sizes.  No clear effect of
Hubble type or distance (resolution) is observed.

\section{PROPERTIES OF THE OB ASSOCIATIONS}

\subsection{The stellar luminosity function\label{LF}}
Many studies of extragalactic stellar populations use the differential
luminosity function (dLF) to investigate possible differences in the
properties of the most massive stars (\cite{freedman85},
\cite{blaha89}). We have determined
the LF in \V\ for all stars contained in the 79 associations, as well
as for all stars in our frames, regardless of their position, and for
field stars only.  We followed Freedman (1985) and
\cite{berkhuijsen89} in using different color selection criteria to
better isolate the blue stars.  It was pointed out by Freedman (1985)
that a \UV\ criterion is to be preferred over \BV\, because the former
is a better discriminator against A supergiant stars.  Adopting the
usual power-law expression for the LF we then determined the slope
$\alpha = d\,$log$N$/$d\,V$ for different color cut-offs, as given by
a least-square fit and using 0.5 mag interval bins.  The color
criterion that provides the largest slope is the adopted one.
Fig.~\ref{lfv.com} shows the LF in \V\ for stars with \UV $\le -0.5$.
The break in the function for \V$>$23.5 suggests that the data become
incomplete at that magnitude, and hence we only include stars with
\V$<$23.5 in our fit.  Table~\ref{lf.table} lists the best fitted power-law 
slopes for various color criteria, for the entire field population and
stars in associations only.  The errors listed include statistical
uncertainties but not systematic effects due to crowding or
incompleteness. 
There is no important difference among the three
color criteria.  The slightly smaller value from the \BV\ selection is
hardly significant, and, if real, might be due to the inclusion of
some later-type stars.  It is interesting however that no difference
is seen between stars in associations and stars in the field.  We
conclude that the slope of the \V\ LF in this field is 
$\sim0.60\pm0.05$. 

Freedman (1985) found a value of 0.67 $\pm$ 0.03 for the \V\ LF of the
brightest stars in nearby galaxies. More recent determinations from
HST observations are given, among others, by \cite{hughes94} and
\cite{hunter95}. The values found (0.56--0.58 in the case of M81, and
between 0.58 and 0.65 for different sets of stars in I Zw 18) are
consistent with the slope determined for M101, and confirm that the
upper LF does not change significantly. This is perhaps not surprising
given the relative insensitivity of the LF slope to changes in the IMF
(\cite{massey85}). However these results appear to rule out any
radical variation in IMF slope or stellar mass limit (see
\S~\ref{imf}).

\subsection{Reddening\label{reddening}}

The substantial extinction in this M101 field is readily inferred from
the optical images, which show many dust filaments along the spiral
arms, and by the color (\BV~$\simeq$~0) of the blue plume in the CMD
of Fig.~\ref{tracks.com}. Hence it is important to measure the
reddening of the OB associations.  The extinction law in M101 was
found to be similar to the Galactic one (R$_V$~=~3.16,
R$_V$~=~A$_V$/E(\BV)) by
\cite{rosa94}.  The foreground extinction in the direction of M101 is
virtually zero (\cite{RC3}).

To measure the reddenning we used the Johnson $Q$ parameter technique,
where $Q$~=~(\UB)$-$0.72$\,$(\BV) is a reddening-free quantity. The
$Q$ value for stars in associations with \BV $<$ 0.1 was used to
derive, by means of the equations given in \cite{massey95}, an
intrinsic color, (\BV)$_0$, which allowed individual stellar
reddenings to be calculated. The mean value for all stars belonging to
the same association was then adopted as the reddening for the entire
group. The estimated uncertainty is
\ebv\ = $\pm$0.1. For the stars in our associations the average
$<$E(\BV)$>$~=~0.21. \cite{wilson91} gives an average of 0.3 mag for
associations in the inner region of M33 (0.15 in an outer region
studied by Regan and Wilson 1993), and values in the range 0.2$-$0.4
are reported, for example, by Haiman \etal (1994).

The ground-based narrow-band imaging was used to measure the
extinction for a number of H~II regions, some of which match the
position of an OB association. The reddening can be measured by the
Balmer decrement: $$\rmn
\frac{I_{H\alpha}}{I_{H\beta}}=\frac{F_{H\alpha}}{F_{H\beta}}10^{C(H\beta)
\cdot {\em f}(\lambda)}$$ where we assume the theoretical value $\rmn
I_{H\alpha}/I_{H\beta}=2.86$ (case B, T=10$^4$ K, N$_e$=100 cm$^{-3}$,
\cite{osterbrock89}) and the extinction curve $f\,$($\lambda$) of
\cite{seaton79}.  The visual extinction is given as
A$_V=2.15\cdot$C(H$\beta$) (Rosa and Benvenuti 1994). The average
reddening for 35 H~II regions is 0.39 mag, with a large scatter (0.19
mag). Of these, 13 coincide with OB associations (nearly 40 percent of
the associations have H~II region counterparts, but these are often
too faint to be included in the analysis).  The comparison with the
values from the broad-band photometry is shown in
Figure~\ref{red.com}.  The Balmer ratio gives, on average, a reddening
value $\sim$0.1 mag larger than the $Q$ method. This is of the same
order as the uncertainties, therefore the difference in the two
distributions is to be considered marginally significant. Note also
that many of the larger reddening values are found for H~II regions
that have no OB association counterpart. We could say that the
identified OB associations do not lie in the regions with the higher
reddening. With such a small sample, though, it is difficult to assert
the reality of this effect. In general we expect however that H~II
regions will be more reddened than the average OB association.

\subsection{The color-magnitude and color-color diagrams} 

We show in Figure~\ref{cmob.com} the CMD of those stars that lie within
the OB associations boundaries. For each association the stars have been
dereddened according to the average \ebv. The CMD morphology is
similar to that of Figure~\ref{tracks.com}, and shows that the selected
OB association boundaries contain several evolved stars. The scatter
of stars across the diagram can be therefore attributed partly to a
spread in age.  The age analysis (next section) tends to confirm the
presence of stellar groups older than 10 Myr among the selected
associations.

The dereddened color-color plot of stars in associations with
photometric uncertainties $<$0.15 mag is shown in
Figure~\ref{ubv.com}. The sequences for dwarf and supergiant stars
have been drawn, adopting the calibrations of \cite{fitzpatrick90} and
\cite{fitzgerald70}. O--B5 stars (\UV $\le -0.9$) are clumped around
the tip of the sequences. A few B5--B9 and A stars are also present at
redder colors. According to the calibration of \cite{fitzpatrick90}
stars of luminosity class Ib have M$_V\simeq-5.2$ ($-$4.3) at \UV$\sim
-0.9$ ($-$0.4) (Ia supergiants are at least one magnitude brighter),
corresponding to $V\simeq24$ (25). At this magnitude both \B\  and \U\
frames are severely incomplete, thus explaining the rapid decline in
the number of stars visible below the \UV$=-$0.9 line.


\subsection{Estimating the age} 

We attempted to estimate the age of the OB associations in M101 by
comparing their dereddened CMDs by visual comparison with theoretical
isochrones (Schaller \etal 1992, Meynet \etal 1993). Usually the
brightest stars were used in the comparison, since the fainter ones
tend to have large photometric errors.  The major sources of
uncertainty in assigning these ages are the reddening, the small
number of stars available in individual associations and photometric
errors.  The presence of binary stars and unresolved clumps is a
further problem.  An uncertainty of $\pm$2 Myr is estimated, based on the
fact that typical photometric and reddening errors can produce a
variation of a few ($\sim 2$) Myr in the calculated ages.

Fig.~\ref{4.com}(a) shows an example of this procedure. The isochrones,
calculated from 2.5 to 9.5 Myr in steps of 1 Myr, are superposed on
the CMD of association \# 37. In this case an age of 4$\pm$2 Myr was
estimated. The evolved stars in the red part of the diagram probably
do not belong to the same episode of star formation that created the
younger stars.  The coexistence of red supergiants and younger OB
stars, pointing to non-coevality, has been noted in previous studies
(\cite{doom85}, \cite{massey89}, Garmany and Stencel 1992). If the
observed red stars are indeed members of the associations, it would
mean that star formation occurs in episodes separated by several Myr,
and that no single age can be assigned.  At least
qualitatively, though, there is no apparent clumping of the red stars
within the associations boundaries. We conclude that they could appear
to belong to the associations simply because of projection effects.  

We show in the other panels of Fig.~\ref{4.com} the dependence on age
of three quantities: the number of blue stars, the size of the
associations and the number of ionizing photons Q$_0$ (calculated from
the H$\alpha$ flux) for the H~II regions which have been identified
also as OB associations.  The estimated uncertainty of $\pm 2$ Myr is
represented by bars elongated along the time axis.  Despite the
scatter, this shows how the richest and largest associations are found
among the youngest ones.  This is probably due to the disruption of
the gravitationally unbound associations with time, combined with
selection effects, which make it easier to detect stars belonging to
young associations.  There is also a dependence of Q$_0$ on age, due
to the relation between the number of blue, ionizing stars and age.
This plot is consistent with the typical H~II region lifetime of 5--6
Myr.  Table~\ref{obdata.table} summarizes the associations properties
presented in this section.

\section{ON THE UPPER MASS LIMIT OF THE IMF\label{imf}}

The use of broad-band photometry to study the IMF of distant
populations of stars is subject to several difficulties. First of all
is the well-known insensitivity of broad-band colors to the effective
temperature of hot, massive stars (Massey \etal 1995a).  This makes
the task of discriminating stars of different masses practically
impossible with photometry alone. Moreover incompleteness in the data
hinders the detection of the hottest stars, whose large bolometric
corrections make them visually fainter than supergiants of later
type. Photometric uncertainties and reddening, together with crowding
and blending of unresolved stellar images (particularly in the young
clusters within the star-forming regions) complicate matters
further.  With these caveats in mind, we can look at the CMDs of
Fig.~\ref{tracks.com} and \ref{cmob.com}. These suggest the presence
of stars more massive than 25--30~M$_\odot$, perhaps 60~M$_\odot$
or higher. The degeneracy of massive stars in this region of the CMD,
though, makes it impossible to say whether very massive stars are
unambiguously present or not.

We also attempted to use an evolutionary synthesis model to constrain
the upper mass limit of the IMF. Integrated magnitudes and colors for
the richest associations (those having more than 15 stars) were
measured by adding up all the flux within the association boundaries.
The reddening-corrected colors were compared with the models of
\cite{leitherer95} (instantaneous burst, solar metallicity). In
Fig.~\ref{models.com} the model predictions for two upper mass
limits, 30 and 100~M$_\odot$, of a power-law IMF with Salpeter's slope
are superposed on the observed colors and ages. While the comparison
for (\UB)$_0$ favors an upper mass limit of 100~M$_\odot$, the other
two colors are however more difficult to interpret.

We conclude from this two exercises that there is some evidence for an
IMF in M101 with a high cut-off mass (comparable to the $\sim$100
$M_\odot$ limit found in the Magellanic Clouds and in the Milky Way by
\cite{massey95b}).  This field has an oxygen abundance of $\sim$1--2
Z$_\odot$ (Kennicutt and Garnett 1996) and some authors have suggested
previously that the H~II regions in this part of M101 have a much
lower upper stellar mass limit (\eg\ \cite{shields76}). Our
observations suggest that such a strong change in the upper mass limit
may not be present, but further observation are needed to make a
definitive test.  This result also agrees with the study of
\cite{rosa94}, who found no need to invoke a metallicity effect on the
IMF in their study of four giant H~II regions in M101.

\section{POPULOUS STAR CLUSTERS}

The importance of star clusters for the understanding of star formation
and evolution has been stressed many times (\cite{vdb91}). The
clusters of the Magellanic Clouds in particular have been the subject
of intense study, leading to a picture of
differing evolutionary histories between the Clouds and the Galaxy. Of
great interest is the presence in the Clouds of populous clusters
(``blue globulars''), which are absent in the Milky Way. Populous
clusters have been observed in a handful of galaxies, and 
\cite{kennicutt-chu} suggested a Hubble type
dependence,
being these clusters preferentially found in
late-type (Sc, Irr) galaxies.  It is therefore interesting to extend the search
for populous clusters to as many galaxies as possible. This could
lead to a better understanding of the properties and systematics of
star formation in galaxies. Observations with HST like those used in
the present work are well suited for this search. With typical sizes
larger than 10 pc in the LMC (\cite{vdb91}), we expect to be able to
identify these objects at least to a distance of 10-12 Mpc on WFPC2
images.

A total of 41 clusters were found by visual search in the WFPC
images, with typical FWHM$ = 0\farcs3$. Their luminosity was measured
with aperture photometry on average images. The estimated uncertainty
in the colors is 0.1 magnitudes, based on the variation in the
measured fluxes using different apertures and sky annuli. The
magnitudes are more uncertain: aperture radii of eight pixels (= 29 pc
in the three Wide Field chips) were adopted, but in some cases smaller
values had to be used to avoid contamination from nearby objects. The
results are summarized in Table~\ref{clusters.table}, while
Fig.~\ref{clust_col_mag.com} and \ref{clust_col_col.com} show the
integrated color-magnitude and color-color diagrams, respectively.

Most clusters are blue (\bv$<$0.5), and occupy the same region in the
(\BV) vs. \V\ diagram as the LMC blue clusters
(Fig.~\ref{clust_col_mag.com}).  The bluest objects are young nuclei
of OB associations, which were included even though they differ from
the stable open clusters.  The color histogram shows a lack of red
clusters relative to the LMC.  This color distribution differs even
more strongly from the colors of clusters in M33 (Christian and
Schommer 1982, 1988) and, especially, in M31 (\cite{hodgemateo87}),
where a larger fraction of clusters have (\bv) $>$ 0.5.

In Fig.~\ref{clust_col_col.com}(a) the sequence of LMC clusters used for
age calibration by \cite{girardi} is indicated.  Using their age
calibration the clusters for which we could measure both \UB\ and \BV\
have ages between a few Myr and $\sim$500 Myr. No reddening correction
has been applied, even though \ebv\ $\simeq 0.1$ could probably bring
the data to a better fit to the LMC sequence.  It seems appropriate to
compare the M101 clusters with the cluster system of M33, also an Sc
galaxy, studied in detail by \cite{christian88}.
Fig.~\ref{clust_col_col.com}(b) compares the (\bv) vs. (\VI) diagram for
the two galaxies.  The calibration line of \cite{christian88} is
shown. The lack of M101 red clusters, having (\VI) $>$ 0.7, is
evident. It could be possible that this is an effect of the position
in the galaxy where these clusters are found, {\it i.e.} close to the
nucleus. In the M33 data, however, there is no indication that the red
clusters preferentially lie away from the nucleus.  The fractional area
of M101 surveyed for clusters is too small to draw firm conclusions on
the overall cluster population and on its differences relative to
other galaxies. We however remind the reader that this area roughly
equals the extent of the entire LMC. On the basis of our analysis we
can only conclude that red clusters (``old globulars'' candidates) are
rare in the field studied. Many of the remaining clusters are very
likely populous clusters of the same kind found in the LMC. The large
ratio between the number of blue and red clusters is consistent with
previous findings that populous clusters are preferentially found in
late-type galaxies. 

Two giant H~II regions, numbers 972 and 1013 in the catalog of
\cite{hodge90}, fall in our HST field, corresponding to associations
57 and 72, respectively. Both have a high H$\alpha$ luminosity, with a
number of ionizing photons in excess of $10^{51}$ s$^{-1}$
(Table~\ref{halpha.table}).  By comparison, 30 Dor in the LMC has an
H$\alpha$ luminosity corresponding to $\sim$10$^{52}$ ionizing
photons/s. 
The morphology of the two regions differs somewhat. Region 72 
resembles 30 Dor in
having a bright, compact core cluster, which is probably responsible
for most of the ionizing flux in the nebula.  Several fainter clusters
or single stars surround the central object. In region 57, on the
contrary, we see a normal association of bright stars, similar in
structure to the giant H~II region NGC~604 in M33.
To better quantify the properties of these embedded star clusters, and
in order to compare them with similar objects in nearby galaxies, we
measured fluxes in different apertures, centered on the brightest
object in each of the two H~II regions.  In region 72, the radius of
the core cluster is $\sim$22 pc, with a radius containing one-half the
total light R$_{0.5}$ $\simeq$ 7 pc, and a total absolute magnitude
M$_V=-12.3$. This object shows some finer structure, namely the
presence of 2 distinct clusters, with a peak ratio of about 3:1. Each
one has R$_{0.5}$ $\simeq$ 3.6 pc. The brightest component has a
luminosity M$_V\simeq -12.0$, and a corresponding mean surface
brightness $\Sigma_{0.5} = 6.9\times10^4 \; L_{V, \odot}$ pc$^{-2}$
inside the radius R$_{0.5}$. This compares with $\Sigma_{0.5} =
1.3\times10^5 \;  L_{V, \odot}$ pc$^{-2}$, R$_{0.5} = 1.7$ pc and a total absolute magnitude $M_V
= -11.1$ for R136 in 30 Dor (\cite{hunter95b}). Region 72 is therefore
comparable to 30 Dor also quantitatively, even though it is somewhat
less compact.  We remind the reader that the ``super star clusters''
found in some galaxies (\cite{oconnell94}) represent more extreme
modes of star formation, 10--30 times more luminous than region 72,
relative to the same age. In region 57 the main component is not as
bright as in region 72, $M_V = -10.7$, while the nearby objects are
typically 2 magnitudes fainter, and could be smaller clusters or very
bright stars. 

\section{Conclusions}
We have described an objective algorithm to identify 79 OB associations
in an HST field of the galaxy M101. The following results were
found:

\noindent
1. The size distribution of the associations is comparable to that
in the Magellanic Clouds, M33, NGC~6822 and M31, with a typical mean size of 90 pc;

\noindent
2. The stellar luminosity function has a slope $d\,$log$N$/$d\,V 
= 0.60 \pm 0.05$, both in the associations and in the general field;

\noindent
3. H~II regions tend to be slightly redder than the average OB
association;

\noindent
4. No indication that the upper mass limit of the IMF is lower
than in low-metallicity environments is found;

\noindent
5. Most of the star clusters identified in the field belong to the
same class of populous clusters found in the LMC. Red clusters are rare.

\acknowledgments 
We thank the HST Distance Scale Key Project team members, Abi Saha in
particular, for their support in this work, Paul Hodge for his
comments and Paul Scowen for providing us with
unpublished material on H~II regions. This work was supported by NASA
through grant GO-2227-87A, and by the NSF through grants AST-9019150
and AST-9421145.

\appendix 

\section{OB ASSOCIATIONS IN THE LMC} 
As a further application of the objective algorithm described in
\S~\ref{identification}, data on the LMC were analyzed.  Stellar
associations in the LMC were catalogued from wide-field photographic
plates by \cite{lucke70} and the characteristics of the 122
associations in the catalog were studied by \cite{hodge70}.  We
applied the clustering algorithm to the catalog of bright stars of
\cite{rousseau78} (in the updated machine-readable form available
through the NASA Astronomical Data Center). It is interesting to
compare our numerical association-finding technique with subjective
human intelligence when applied to the nearest galaxy, where a
resolution of less than a pc (1\arcsec = 0.24 pc) is attainable.

Stars were selected with two different criteria, based on spectral
type (all catalogued stars with type earlier than B2.5) and on
photometric parameters (\V $<$ 13.7, \BV $<$ 0.15, to match the
criteria adopted for M101 and M33).  The two sets of stars gave
approximately the same results, so the following discussion
concentrates on the photometrically selected one.  The search radius
determined by the algorithm is 60 pc (the same used by Battinelli
(1991) in the SMC).  Out of the 1354 ``blue'' stars, 725 are
distributed in 121 associations.  The resulting average size is 78 pc,
the same value found by \cite{hodge70}.  This includes all agglomerations of
stars with at least 3 stars.  A correction, based on the
expected contamination by statistical fluctuations, gives 94 pc
instead.  The distribution of association sizes is shown in
Fig.~\ref{hist.com-lmc} for both Hodge and Lucke's data and this work.
They appear remarkably similar, except for a larger number of small
associations in Hodge and Lucke.  Regarding the case-by-case
comparison (Fig.~\ref{clust.com-lmc}), the agreement is far from good.
It has been noted (\cite{caplan86}) that the stars in the Rousseau
\etal catalog, which is based on an objective prism spectral survey, show no 
concentration towards H~II regions, and we infer that the catalog is
under-representing the regions richest in OB stars, which may be the
explanation for the poor agreement.  For the purpose of making an
objective catalog of associations in the LMC a better catalog of the
brightest blue stars is needed.

\clearpage

\def\V{$V$}

\begin{deluxetable}{ccccc}
\tablenum{1}
\tablewidth{0pt}
\tablecaption{Median photometric errors.\label{errors.table}}
\tablehead{
\colhead{\V\phantom{aaaa}}           & \colhead{$\sigma_{\scriptscriptstyle V}$\phantom{aaa}}      &
\colhead{$\sigma_{\scriptscriptstyle I}$\phantom{aaa}}  &
\colhead{$\sigma_{\scriptscriptstyle B}$\phantom{aaa}}	& \colhead{$\sigma_{\scriptscriptstyle U}$\phantom{aaa}}}

\startdata

20--21\phantom{aaaa}	& 0.031\phantom{aaa}	& 0.044\phantom{aaa}	& 0.094\phantom{aaa}	& 0.115\phantom{aaa} \nl
21--22\phantom{aaaa}	& 0.034\phantom{aaa} & 0.048\phantom{aaa} & 0.117\phantom{aaa} & 0.149\phantom{aaa} \nl
22--23\phantom{aaaa}	& 0.038\phantom{aaa} & 0.062\phantom{aaa} & 0.125\phantom{aaa} & 0.157\phantom{aaa} \nl
23--24\phantom{aaaa}	& 0.048\phantom{aaa} & 0.085\phantom{aaa} & 0.165\phantom{aaa} & 0.202\phantom{aaa} \nl
24--25\phantom{aaaa}	& 0.061\phantom{aaa} & 0.122\phantom{aaa} & 0.235\phantom{aaa} & 0.285\phantom{aaa} \nl
25--26\phantom{aaaa}	& 0.080\phantom{aaa} & 0.181\phantom{aaa} & 0.361\phantom{aaa} & 0.432\phantom{aaa} \nl
26--27\phantom{aaaa}	& 0.107\phantom{aaa} & 0.242\phantom{aaa} & 0.529\phantom{aaa} & 0.618\phantom{aaa} \nl

\enddata
\end{deluxetable}

\def\UBVI{\hbox{$U\!BV\!I$}}
\def\BV{\hbox{$B\!-\!V$}}
\def\UB{\hbox{$U\!-\!B$}}
\def\UV{\hbox{$U\!-\!V$}}
\def\VI{\hbox{$V\!-\!I$}}

\begin{deluxetable}{cccc}
\tablenum{2}
\tablewidth{0pt}
\tablecaption{Slope of the luminosity function.\label{lf.table}}
\tablehead{
\colhead{Color criterion\phantom{aaa}}           & \colhead{field + association\
s}      &
\colhead{field\phantom{aa}}  &
\colhead{associations}  }
\startdata
 
\UV\ $<$ $-$0.5\phantom{aaa} & 0.58 $\pm$ 0.04  & 0.58 $\pm$ 0.05\phantom{aa}  \
& 0.58 $\pm$ 0.06 \nl
\UB\ $<$ $-$0.7\phantom{aaa} & 0.62 $\pm$ 0.04  & 0.62 $\pm$ 0.05\phantom{aa}  \
& 0.62 $\pm$ 0.07 \nl
\BV\ $<$ $+$0.4\phantom{aaa} & 0.55 $\pm$ 0.04  & 0.57 $\pm$ 0.04\phantom{aa}  \
& 0.53 $\pm$ 0.05 \nl
 
\enddata
\end{deluxetable}


\def\rmn{\everymath={\fam0 }\fam0 } \def\hii{H~II}




\begin{deluxetable}{cccccccc}
\tablenum{3}

\tablewidth{0pt}
\tablecaption{Properties of OB associations.\label{obdata.table}}
\tablehead{
\colhead{ID}           & \colhead{RA ($\rmn 14^h$)}      &
\colhead{DEC (+54\arcdeg)}  &
\colhead{Diameter (pc)}    &
\colhead{$\rmn N_{blue}$}  & \colhead{$\rmn N_{tot}$}  &
\colhead{E(B$-$V)} & \colhead{age (Myr)}}

\startdata

    1   &    $\rmn 3^m$    25\fs1   &   21\arcmin\/      10\arcsec   &21   &   11   &   15   &    0   &   14 \nl
    2   &    $\rmn 3^m$    23\fs2   &   21\arcmin\/      25\arcsec   &103   &   12   &   35   &  0.3   &    5 \nl
    3   &    $\rmn 3^m$    24\fs4   &   21\arcmin\/      05\arcsec   &29   &    6   &   12   &  0.3   &    5 \nl
    4   &    $\rmn 3^m$    24\fs1   &   21\arcmin\/      02\arcsec   &61   &   16   &   45   &  0.4   &    4 \nl
    5   &    $\rmn 3^m$    24\fs2   &   20\arcmin\/      55\arcsec   &72   &   14   &   26   &  0.4   &    6 \nl
    6   &    $\rmn 3^m$    23\fs9   &   21\arcmin\/      00\arcsec   &118   &   11   &   29   &  0.3   &    3 \nl
    7   &    $\rmn 3^m$    22\fs4   &   21\arcmin\/      11\arcsec   &62   &    6   &    8   &  0.2   &    6 \nl
    8   &    $\rmn 3^m$    17\fs6   &   21\arcmin\/      00\arcsec   &119   &   10   &   25   &  0.2   &    8 \nl
    9   &    $\rmn 3^m$    17\fs9   &   21\arcmin\/      06\arcsec   &306   &   49   &  125   &  0.2   &    5 \nl
   10   &    $\rmn 3^m$    18\fs8   &   21\arcmin\/      23\arcsec   &126   &   12   &   24   &  0.3   &    5 \nl
   11   &    $\rmn 3^m$    21\fs6   &   21\arcmin\/      45\arcsec   &96   &   10   &   18   &  0.2   &    4 \nl
   12   &    $\rmn 3^m$    16\fs6   &   21\arcmin\/      13\arcsec   &80   &   10   &   16   &  0.2   &    5 \nl
   13   &    $\rmn 3^m$    19\fs7   &   21\arcmin\/      44\arcsec   &89   &   11   &   20   &  0.2   &    4 \nl
   14   &    $\rmn 3^m$    15\fs8   &   21\arcmin\/      24\arcsec   &108   &    8   &   15   &  0.2   &    4 \nl
   15   &    $\rmn 3^m$    17\fs8   &   21\arcmin\/      42\arcsec   &113   &   14   &   26   &  0.2   &    6 \nl
   16   &    $\rmn 3^m$    17\fs0   &   21\arcmin\/      41\arcsec   &127   &   12   &   28   &  0.3   &    4 \nl
   17   &    $\rmn 3^m$    16\fs0   &   21\arcmin\/      45\arcsec   &129   &   24   &   39   &  0.1   &    4 \nl
   18   &    $\rmn 3^m$    15\fs7   &   21\arcmin\/      46\arcsec   &154   &    9   &   15   &  0.1   &    6 \nl
   19   &    $\rmn 3^m$    18\fs1   &   22\arcmin\/      04\arcsec   &111   &   10   &   19   &  0.2   &    4 \nl
   20   &    $\rmn 3^m$    15\fs6   &   21\arcmin\/      52\arcsec   &226   &   21   &   63   &  0.2   &    6 \nl
   21   &    $\rmn 3^m$    15\fs0   &   21\arcmin\/      51\arcsec   &43   &   11   &   12   &  0.1   &    4 \nl
   22   &    $\rmn 3^m$    13\fs1   &   21\arcmin\/      50\arcsec   &173   &   15   &   30   &  0.2   &    6 \nl
   23   &    $\rmn 3^m$    18\fs1   &   22\arcmin\/      20\arcsec   &128   &    8   &   12   &  0.1   &    7 \nl
   24   &    $\rmn 3^m$    12\fs6   &   21\arcmin\/      46\arcsec   &121   &   15   &   30   &  0.2   &    7 \nl
   25   &    $\rmn 3^m$    13\fs1   &   21\arcmin\/      53\arcsec   &59   &    7   &    8   &  0.3   &    4 \nl
   26   &    $\rmn 3^m$    12\fs6   &   21\arcmin\/      55\arcsec   &77   &    7   &   10   &  0.1   &    6 \nl
   27   &    $\rmn 3^m$    18\fs4   &   22\arcmin\/      33\arcsec   &143   &   16   &   21   &  0.2   &    7 \nl
   28   &    $\rmn 3^m$    21\fs4   &   22\arcmin\/      42\arcsec   &125   &   10   &   23   &  0.1   &    6 \nl
   29   &    $\rmn 3^m$    24\fs3   &   22\arcmin\/      12\arcsec   &78   &    9   &   15   &  0.2   &    3 \nl
   30   &    $\rmn 3^m$    26\fs3   &   22\arcmin\/      00\arcsec   &105   &    8   &   21   &  0.3   &    6 \nl
   31   &    $\rmn 3^m$    25\fs7   &   22\arcmin\/      14\arcsec   &231   &   30   &   65   &  0.3   &    4 \nl
   32   &    $\rmn 3^m$    26\fs1   &   22\arcmin\/      11\arcsec   &181   &   22   &   40   &  0.2   &    4 \nl
   33   &    $\rmn 3^m$    26\fs8   &   22\arcmin\/      04\arcsec   &102   &   10   &   19   &  0.2   &    7 \nl
   34   &    $\rmn 3^m$    25\fs9   &   22\arcmin\/      26\arcsec   &117   &   10   &   18   &  0.2   &    6 \nl
   35   &    $\rmn 3^m$    23\fs6   &   22\arcmin\/      55\arcsec   &74   &    9   &   11   &  0.2   &   10 \nl
   36   &    $\rmn 3^m$    25\fs4   &   22\arcmin\/      43\arcsec   &162   &   17   &   34   &  0.2   &    6 \nl
   37   &    $\rmn 3^m$    24\fs6   &   22\arcmin\/      55\arcsec   &232   &   27   &  101   &  0.3   &    4 \nl
   38   &    $\rmn 3^m$    28\fs3   &   22\arcmin\/      08\arcsec   &141   &    9   &   20   &  0.4   &    4 \nl
   39   &    $\rmn 3^m$    25\fs5   &   22\arcmin\/      55\arcsec   &83   &    7   &   11   &  0.2   &    4 \nl
   40   &    $\rmn 3^m$    25\fs3   &   21\arcmin\/      16\arcsec   &68   &   10   &   17   &  0.3   &    4 \nl
   41   &    $\rmn 3^m$    24\fs6   &   21\arcmin\/      24\arcsec   &83   &   10   &   15   &  0.1   &    7 \nl
   42   &    $\rmn 3^m$    25\fs2   &   21\arcmin\/      15\arcsec   &59   &    8   &   13   &    0   &   14 \nl
   43   &    $\rmn 3^m$    25\fs6   &   21\arcmin\/      10\arcsec   &80   &    9   &   15   &  0.2   &    6 \nl
   44   &    $\rmn 3^m$    25\fs3   &   21\arcmin\/      12\arcsec   &93   &   10   &   17   &  0.2   &    6 \nl
   45   &    $\rmn 3^m$    25\fs5   &   21\arcmin\/      08\arcsec   &94   &    8   &   16   &  0.2   &    7 \nl
   46   &    $\rmn 3^m$    24\fs4   &   21\arcmin\/      22\arcsec   &66   &   11   &   16   &    0   &   11 \nl
   47   &    $\rmn 3^m$    25\fs7   &   21\arcmin\/      05\arcsec   &51   &    9   &    9   &  0.2   &    4 \nl
   48   &    $\rmn 3^m$    25\fs1   &   21\arcmin\/      13\arcsec   &59   &   10   &   13   &  0.3   &    8 \nl
   49   &    $\rmn 3^m$    25\fs8   &   21\arcmin\/      04\arcsec   &38   &    8   &    9   &  0.1   &    6 \nl
   50   &    $\rmn 3^m$    25\fs6   &   21\arcmin\/      06\arcsec   &80   &    7   &   15   &  0.3   &    7 \nl
   51   &    $\rmn 3^m$    23\fs7   &   21\arcmin\/      22\arcsec   &126   &   23   &   41   &  0.3   &    3 \nl
   52   &    $\rmn 3^m$    23\fs4   &   21\arcmin\/      26\arcsec   &46   &    8   &   11   &  0.3   &    8 \nl
   53   &    $\rmn 3^m$    23\fs5   &   21\arcmin\/      25\arcsec   &49   &    8   &   10   &  0.2   &    4 \nl
   54   &    $\rmn 3^m$    23\fs9   &   21\arcmin\/      19\arcsec   &140   &   19   &   37   &  0.3   &    5 \nl
   55   &    $\rmn 3^m$    23\fs8   &   21\arcmin\/      18\arcsec   &91   &   10   &   15   &  0.2   &    9 \nl
   56   &    $\rmn 3^m$    23\fs7   &   21\arcmin\/      20\arcsec   &59   &    7   &   10   &  0.2   &    5 \nl
   57   &    $\rmn 3^m$    23\fs7   &   21\arcmin\/      16\arcsec   &203   &  105   &  175   &  0.2   &    3 \nl
   58   &    $\rmn 3^m$    24\fs0   &   21\arcmin\/      14\arcsec   &89   &   12   &   21   &  0.2   &    7 \nl
   59   &    $\rmn 3^m$    24\fs1   &   21\arcmin\/      13\arcsec   &53   &    8   &   10   &    0   &    9 \nl
   60   &    $\rmn 3^m$    24\fs2   &   21\arcmin\/      11\arcsec   &53   &   11   &   12   &  0.4   &    4 \nl
   61   &    $\rmn 3^m$    23\fs1   &   21\arcmin\/      21\arcsec   &99   &   23   &   36   &  0.2   &    4 \nl
   62   &    $\rmn 3^m$    23\fs2   &   21\arcmin\/      16\arcsec   &76   &   13   &   31   &  0.3   &    7 \nl
   63   &    $\rmn 3^m$    23\fs3   &   21\arcmin\/      15\arcsec   &133   &   22   &   50   &  0.2   &    6 \nl
   64   &    $\rmn 3^m$    23\fs0   &   21\arcmin\/      18\arcsec   &122   &   22   &   41   &  0.3   &    4 \nl
   65   &    $\rmn 3^m$    23\fs1   &   21\arcmin\/      17\arcsec   &77   &   13   &   14   &  0.2   &    5 \nl
   66   &    $\rmn 3^m$    23\fs2   &   21\arcmin\/      14\arcsec   &79   &    9   &   12   &  0.3   &    7 \nl
   67   &    $\rmn 3^m$    24\fs0   &   21\arcmin\/      05\arcsec   &71   &   12   &   19   &  0.3   &    6 \nl
   68   &    $\rmn 3^m$    22\fs7   &   21\arcmin\/      16\arcsec   &230   &   47   &  116   &  0.3   &    4 \nl
   69   &    $\rmn 3^m$    22\fs6   &   21\arcmin\/      17\arcsec   &74   &    9   &   11   &  0.1   &    7 \nl
   70   &    $\rmn 3^m$    22\fs7   &   21\arcmin\/      08\arcsec   &98   &   11   &   24   &  0.2   &    5 \nl
   71   &    $\rmn 3^m$    22\fs8   &   21\arcmin\/      09\arcsec   &88   &   14   &   23   &  0.1   &    6 \nl
   72   &    $\rmn 3^m$    22\fs8   &   21\arcmin\/      04\arcsec   &241   &   88   &  221   &  0.2   &    3 \nl
   73   &    $\rmn 3^m$    22\fs0   &   21\arcmin\/      09\arcsec   &82   &   10   &   19   &  0.3   &    7 \nl
   74   &    $\rmn 3^m$    21\fs5   &   21\arcmin\/      13\arcsec   &61   &    7   &    9   &  0.2   &    7 \nl
   75   &    $\rmn 3^m$    21\fs6   &   21\arcmin\/      11\arcsec   &65   &    8   &   12   &    0   &   12 \nl
   76   &    $\rmn 3^m$    21\fs4   &   21\arcmin\/      13\arcsec   &59   &    8   &   10   &  0.2   &    7 \nl
   77   &    $\rmn 3^m$    21\fs2   &   21\arcmin\/      15\arcsec   &71   &   15   &   22   &  0.3   &    5 \nl
   78   &    $\rmn 3^m$    21\fs5   &   21\arcmin\/      12\arcsec   &67   &    7   &   12   &  0.2   &   10 \nl
   79   &    $\rmn 3^m$    21\fs1   &   21\arcmin\/      13\arcsec   &68   &    7   &    8   &  0.2   &    9 \nl

\tablenotetext{a}{Diameters are for a distance of 7.4 Mpc}
\enddata
\end{deluxetable}



\begin{deluxetable}{cccc}
\tablenum{4}
\tablewidth{0pt}
\tablecaption{Comparison of OB associations properties.\label{compare.table}}
\tablehead{
\colhead{Galaxy\phantom{aa}}           & \colhead{average}      &
\colhead{median}  &
\colhead{minimum no.} \\
\colhead{}	& \colhead{diameter (pc)}	& \colhead{diameter (pc)}	& \colhead{of stars}  }

\startdata

M 101\phantom{aa}	& 100	& 90	& 7 \nl
LMC\phantom{aa}	& 80	& 60	& 3 \nl
SMC\phantom{aa}	& 90	& 70	& 3 \nl
M 33\phantom{aa}	& 80	& 60	& 10 \nl
M 31\phantom{aa}	& 120	& 100	& 5 \nl
NGC 6822\phantom{aa}        & 90   & 90   & 10 \nl

\enddata
\end{deluxetable}

\begin{deluxetable}{ccc}
\tablenum{5}
\tablewidth{0pt}
\tablecaption{Balmer decrement reddening and number of
ionizing photons.\label{halpha.table}}
\tablehead{
\colhead{Association no.}           & \colhead{E($B-V$)}      &
\colhead{log $Q_0$}}
 
\startdata
 
4       & 0.82  & 50.91 \nl
11      & 0.19  & 49.91 \nl
13      & 0.29  & 50.33 \nl
31      & 0.29  & 50.22 \nl
36      & 0.36  & 49.88 \nl
41      & 0.14  & 49.25 \nl
51      & 0.19  & 49.91 \nl
52      & 0.41  & 49.71 \nl
57      & 0.24  & 51.17 \nl
61      & 0.47  & 50.27 \nl
67      & 0.25  & 50.12 \nl
68      & 0.39  & 50.94 \nl
72      & 0.29  & 51.92 \nl

\enddata
\end{deluxetable}


\def\rmn{\everymath={\fam0 }\fam0 } \def\hii{H~II}

\begin{deluxetable}{ccccccc}
\tablenum{6}
\tablewidth{0pt}
\tablecaption{Cluster photometry.\label{clusters.table}}
\tablehead{
\colhead{ID\phantom{aaa}}           & \colhead{RA ($\rmn 14^h$)}	&
\colhead{DEC (+54\arcdeg)}  &
\colhead{V}\phantom{aa}      &
\colhead{U-B}\phantom{aa}  &
\colhead{B-V}\phantom{aa}    &
\colhead{V-I}\phantom{aa}}

\startdata

1\phantom{aaa}  & $\rmn 3^m$  25\fs3  &   21\arcmin\/  11\arcsec &   20.51\phantom{aa} &   -0.50\phantom{aa} &   0.21\phantom{aa} &   0.55\phantom{aa} \nl
2\phantom{aaa}  & $\rmn 3^m$  25\fs5  &   21\arcmin\/  06\arcsec &   20.93\phantom{aa} &   -0.80\phantom{aa} &   0.26\phantom{aa} &   0.51\phantom{aa} \nl
3\phantom{aaa}  & $\rmn 3^m$  23\fs7  &   21\arcmin\/  14\arcsec &   23.26\phantom{aa} & \nodata\phantom{aa} &   0.58\phantom{aa} &   0.67\phantom{aa} \nl
4\phantom{aaa}  & $\rmn 3^m$  22\fs4  &   21\arcmin\/  05\arcsec &   23.71\phantom{aa} & \nodata\phantom{aa}        &   0.24\phantom{aa} &   0.45\phantom{aa} \nl
5\phantom{aaa}  & $\rmn 3^m$  23\fs4  &   21\arcmin\/  28\arcsec &   21.59\phantom{aa} & \nodata\phantom{aa}        &   0.38\phantom{aa} &   0.60\phantom{aa} \nl
6\phantom{aaa}  & $\rmn 3^m$  22\fs7  &   20\arcmin\/  58\arcsec &   22.48\phantom{aa} & \nodata\phantom{aa}        &   0.30\phantom{aa} &   0.49\phantom{aa} \nl
7\phantom{aaa}  & $\rmn 3^m$  25\fs0  &   21\arcmin\/  12\arcsec &   22.29\phantom{aa} & \nodata\phantom{aa}        &   0.31\phantom{aa} &   0.53\phantom{aa} \nl
8\phantom{aaa}  & $\rmn 3^m$  25\fs3  &   21\arcmin\/  08\arcsec &   22.51\phantom{aa} & \nodata\phantom{aa}        &   0.06\phantom{aa} &   0.39\phantom{aa} \nl
9\phantom{aaa}  & $\rmn 3^m$  24\fs6  &   21\arcmin\/  04\arcsec &   22.07\phantom{aa} &   -0.67\phantom{aa} &   0.29\phantom{aa} &   0.55\phantom{aa} \nl
10\phantom{aaa} & $\rmn 3^m$  19\fs9  &   21\arcmin\/  13\arcsec &   22.31\phantom{aa} &   -0.23\phantom{aa} &   0.10\phantom{aa} &   0.36\phantom{aa} \nl
11\phantom{aaa} & $\rmn 3^m$  16\fs9  &   21\arcmin\/  19\arcsec &   21.88\phantom{aa} &   -0.07\phantom{aa} &   0.33\phantom{aa} &   0.62\phantom{aa} \nl
12\phantom{aaa} & $\rmn 3^m$  18\fs7  &   22\arcmin\/  33\arcsec &   20.86\phantom{aa}&   -0.79\phantom{aa} &  -0.18\phantom{aa} &   0.09\phantom{aa} \nl
13\phantom{aaa} & $\rmn 3^m$  21\fs9  &   21\arcmin\/  45\arcsec &   20.45\phantom{aa} &   -0.89\phantom{aa} &  -0.02\phantom{aa} &   0.34\phantom{aa} \nl
14\phantom{aaa} & $\rmn 3^m$  15\fs6  &   22\arcmin\/  13\arcsec &   21.00\phantom{aa} &    0.34\phantom{aa} &   0.27\phantom{aa} &   0.58\phantom{aa} \nl
15\phantom{aaa} & $\rmn 3^m$  14\fs1  &   22\arcmin\/  05\arcsec &   20.95\phantom{aa} &   -0.27\phantom{aa} &   0.12\phantom{aa} &   0.52\phantom{aa} \nl
16\phantom{aaa} & $\rmn 3^m$  17\fs7  &   22\arcmin\/  09\arcsec &   21.27\phantom{aa} &    0.11\phantom{aa} &   0.25\phantom{aa} &   0.71\phantom{aa} \nl
17\phantom{aaa} & $\rmn 3^m$  19\fs7  &   21\arcmin\/  40\arcsec &   20.04\phantom{aa} & \nodata\phantom{aa}        &   1.01\phantom{aa} &   1.08\phantom{aa} \nl
18\phantom{aaa} & $\rmn 3^m$  18\fs5  &   22\arcmin\/  33\arcsec &   21.65\phantom{aa} &    0.04\phantom{aa} &   0.43\phantom{aa} &   0.59\phantom{aa} \nl
19\phantom{aaa} & $\rmn 3^m$  22\fs2  &   22\arcmin\/  49\arcsec &   21.52\phantom{aa} &   -0.41\phantom{aa} &   0.16\phantom{aa} &   0.60\phantom{aa} \nl
20\phantom{aaa} & $\rmn 3^m$  25\fs6  &   22\arcmin\/  41\arcsec &   21.00\phantom{aa} &   -0.06\phantom{aa} &   0.20\phantom{aa} &   0.37\phantom{aa} \nl
21\phantom{aaa} & $\rmn 3^m$  27\fs4  &   22\arcmin\/  15\arcsec &   21.76\phantom{aa} &   -0.50\phantom{aa} &   0.13\phantom{aa} &   0.68\phantom{aa} \nl
22\phantom{aaa} & $\rmn 3^m$  26\fs4  &   21\arcmin\/  58\arcsec &   21.71\phantom{aa} &   -0.66\phantom{aa} &   0.30\phantom{aa} &   0.99\phantom{aa} \nl
23\phantom{aaa} & $\rmn 3^m$  29\fs7  &   22\arcmin\/  24\arcsec &   20.84\phantom{aa} &    0.09\phantom{aa} &   0.35\phantom{aa} &   0.65\phantom{aa} \nl
24\phantom{aaa} & $\rmn 3^m$  29\fs3  &   22\arcmin\/  19\arcsec &   21.35\phantom{aa} & \nodata\phantom{aa}        &   0.46\phantom{aa} &   0.72\phantom{aa} \nl
25\phantom{aaa} & $\rmn 3^m$  28\fs4  &   22\arcmin\/  19\arcsec &   22.26\phantom{aa} & \nodata\phantom{aa}        &   0.39\phantom{aa} &   0.66\phantom{aa} \nl
26\phantom{aaa} & $\rmn 3^m$  22\fs2  &   22\arcmin\/  37\arcsec &   21.74\phantom{aa} & \nodata\phantom{aa}        &   0.46\phantom{aa} &   0.97\phantom{aa} \nl
27\phantom{aaa} & $\rmn 3^m$  24\fs1  &   21\arcmin\/  43\arcsec &   21.85\phantom{aa} &   -0.93\phantom{aa} &   -0.07\phantom{aa} &   0.48\phantom{aa} \nl 
28\phantom{aaa} & $\rmn 3^m$  24\fs1  &   21\arcmin\/  57\arcsec &   22.10\phantom{aa} &   -1.00\phantom{aa} &   0.05\phantom{aa} &   0.29\phantom{aa} \nl
29\phantom{aaa} & $\rmn 3^m$  25\fs7  &   22\arcmin\/  09\arcsec &   21.53\phantom{aa} &   -0.97\phantom{aa} &  -0.02\phantom{aa} &   0.31\phantom{aa} \nl
30\phantom{aaa} & $\rmn 3^m$  24\fs2  &   20\arcmin\/  56\arcsec &   20.50\phantom{aa} &   -0.06\phantom{aa} &   0.22\phantom{aa} &   0.55\phantom{aa} \nl
31\phantom{aaa} & $\rmn 3^m$  24\fs2  &   21\arcmin\/  10\arcsec &   20.46\phantom{aa} &   -0.61\phantom{aa} &   0.29\phantom{aa} &   0.69\phantom{aa} \nl
32\phantom{aaa} & $\rmn 3^m$  24\fs0  &   21\arcmin\/  12\arcsec &   20.56\phantom{aa} &   -0.51\phantom{aa} &   0.38\phantom{aa} &   0.59\phantom{aa} \nl
33\phantom{aaa} & $\rmn 3^m$  22\fs3  &   21\arcmin\/  12\arcsec &   22.34\phantom{aa} & \nodata\phantom{aa}        &   0.99\phantom{aa} &   1.01\phantom{aa} \nl
34\phantom{aaa} & $\rmn 3^m$  24\fs0  &   21\arcmin\/  23\arcsec &   19.58\phantom{aa} & \nodata\phantom{aa}        &   1.14\phantom{aa} &   1.47\phantom{aa} \nl
35\phantom{aaa} & $\rmn 3^m$  21\fs8  &   21\arcmin\/  05\arcsec &   20.76\phantom{aa} &    0.33\phantom{aa} &   0.44\phantom{aa} &   0.70\phantom{aa} \nl
36\phantom{aaa} & $\rmn 3^m$  24\fs8  &   21\arcmin\/  11\arcsec &   21.49\phantom{aa} &    0.17\phantom{aa} &   0.78\phantom{aa} &   1.19\phantom{aa} \nl
37\phantom{aaa} & $\rmn 3^m$  24\fs6  &   21\arcmin\/  11\arcsec &   21.59\phantom{aa} &    0.01\phantom{aa}  &   0.27\phantom{aa} &   0.54\phantom{aa} \nl
38\phantom{aaa} & $\rmn 3^m$  24\fs5  &   21\arcmin\/  02\arcsec &   21.48\phantom{aa} &   -0.35\phantom{aa}  &   0.38\phantom{aa} &   0.60\phantom{aa} \nl
39\phantom{aaa} & $\rmn 3^m$  24\fs9  &   21\arcmin\/  11\arcsec &   21.91\phantom{aa}  &   -0.47\phantom{aa}  &   0.42\phantom{aa} &   0.69\phantom{aa} \nl
40\phantom{aaa} & $\rmn 3^m$  23\fs3  &   20\arcmin\/  53\arcsec &   21.46\phantom{aa} &   -0.42\phantom{aa}  &   0.29\phantom{aa} &   0.55\phantom{aa} \nl
41\phantom{aaa} & $\rmn 3^m$  23\fs2  &   21\arcmin\/  01\arcsec &   22.03\phantom{aa}  & \nodata\phantom{aa}        &   0.51\phantom{aa} &   0.82\phantom{aa} \nl
42\tablenotemark{1}\phantom{aaa} &  $\rmn 3^m$  22\fs8  & 21\arcmin\/ 04\arcsec & 17.71\phantom{aa} & -1.10\phantom{aa} & 0.08\phantom{aa} &   0.05\phantom{aa} \nl
43\tablenotemark{2}\phantom{aaa} &  $\rmn 3^m$  23\fs7  & 21\arcmin\/ 16\arcsec & 19.38\phantom{aa} & -1.05\phantom{aa} & 0.25\phantom{aa} &  -0.09\phantom{aa} \nl

\enddata
\tablenotetext{1}{H~II region core cluster, in association~72}
\tablenotetext{2}{H~II region core cluster, in association 57}
\end{deluxetable}







\clearpage

\figcaption[Bresolin.fig1.ps]{Mosaic of the four WFPC2 chips. H~II
regions 1013 and 972 from the catalog of Hodge \etal (1990)
are the brightest objects in chip 4 (upper right)(courtesy
A. Turner).\label{mosaic}}

\figcaption[Bresolin.fig2.ps]{H$\alpha$ (line + continuum) image of M101
showing the HST field studied in this work. North is at the
top.\label{halphaimage}}

\figcaption[Bresolin.fig3.ps]{Color-magnitude diagram of the M101 field with
evolutionary tracks from Schaller \etal (1992),
reddened by 0.21 mag in (\BV). {\it (Inset)} The stars brighter than \V$=22$ 
that appear in relatively uncrowded regions are shown with their
individual photometric errors.\label{tracks.com}}

\figcaption[Bresolin.fig4.ps]{Outlines of the 79 OB associations which were
found by applying the objective clustering algorithm.
Stars having (\BV) $<$ 0.4 and \V\ $<$ 25.5 are plotted.
\label{clust.com}}

\figcaption[Bresolin.fig5.ps]{Comparison of associations size distribution for
M101, LMC, SMC, M31, NGC~6822 and M33.\label{size.com}}

\figcaption[Bresolin.fig6.ps]{The differential \V\ luminosity function for the
brightest stars in the associations. Only stars having (\UV)
$<-0.5$ are included.\label{lfv.com}}

\figcaption[Bresolin.fig7.ps]{(a) \ebv\ for all OB associations
(measured by the reddening-free parameter
$Q$=(\UB)$-0.72\,$(\BV)) and for all the bright H~II regions
found in the M101 field (measured by the Balmer decrement). 
(b) \ebv\ for those
associations for which both $Q$ and the Balmer decrement were
measured.\label{red.com}}

\figcaption[Bresolin.fig8.ps]{Color-magnitude diagram for the stars within the
OB associations boundaries. For each association the stars
have been dereddened according to the average \ebv\ value
measured from the $Q$ parameter.\label{cmob.com}}

\figcaption[Bresolin.fig9.ps]{(\UB) {\it vs} (\BV) diagram for those stars in
associations having internal photometric errors smaller than
0.15 mag in each band. The reddening correction has been
applied as in Fig.~\ref{cmob.com}. The main sequence and supergiants
sequence are plotted, together with two lines of constant
\UV\ color ($-0.9$, supergiants of spectral type B5, and
$-0.4$, at the transition between late B and early A
supergiants). The arrow represents the reddening vector for
a B0 supergiant.\label{ubv.com}}

\figcaption[Bresolin.fig10.ps]{(a) Example of age determination: theoretical
isochrones from the Schaller \etal (1992)
models from 2.5 to 9.5
Myr in steps of 1 Myr are superposed on 
the dereddened CMD of association 37. 
In this case an age of 4
Myr ($\pm$2 Myr)
was estimated. (b) Number of blue stars in the
associations vs. age.  (c) Diameter of the associations
vs. age. (d) Number of ionizing photons (measured from
the H$\alpha$ flux) vs. age. In these plots the
estimated uncertainty of $\pm$2 Myr is represented by
elongated bars.\label{4.com}}

\figcaption[Bresolin.fig11.ps]{The integrated and dereddened colors of the
richest associations (N$_{\rmn blue}>15$) compared to Leitherer and
Heckman's (1995) population models for an instantaneous burst at solar
metallicity.  The dots represent the observed colors and the assigned
ages.  The two lines are for a Salpeter IMF with high-mass limits of
100 and 30 M$_\odot$.\label{models.com}}

\figcaption[Bresolin.fig12.ps]{(a) (\BV) vs. \V\ integrated color-magnitude diagram 
for star clusters in M101 (open
circles) and in the LMC (dots, from van den Bergh 1981); (b)
color distribution for M101 (hatched) and LMC clusters.
\label{clust_col_mag.com}}

\figcaption[Bresolin.fig13.ps]{(a) (\BV) vs. (\UB) 
integrated color-color diagram for the M101 clusters. The sequence of
LMC clusters from Girardi \etal (1995) is indicated; (b) (\VI)
vs. (\BV) diagram for clusters in M101 (open circles) and M33
(dots). The theoretical sequence shown is taken from Christian and
Schommer (1988).\label{clust_col_col.com}}

\figcaption[Bresolin.fig14.ps]{Histogram of the sizes of the associations in the LMC
as determined by the objective algorithm (continuous line) and those
found by Lucke and Hodge (dashed line). In the latter case more small
agglomerations were found, and less for sizes of 80-100 pc. In
general, though, the two distributions are very
similar.\label{hist.com-lmc}}

\figcaption[Bresolin.fig15.ps]{This map shows the LMC association boundaries
determined by the objective algorithm (irregular polygons)
together with the associations in the Lucke and Hodge (1970)
catalog (circles). Although there are several matching
cases, many discrepancies are present. This is likely to be
due to the incompleteness of the bright LMC stars catalog in
the densest regions and in correspondence of H~II regions.
\label{clust.com-lmc}}

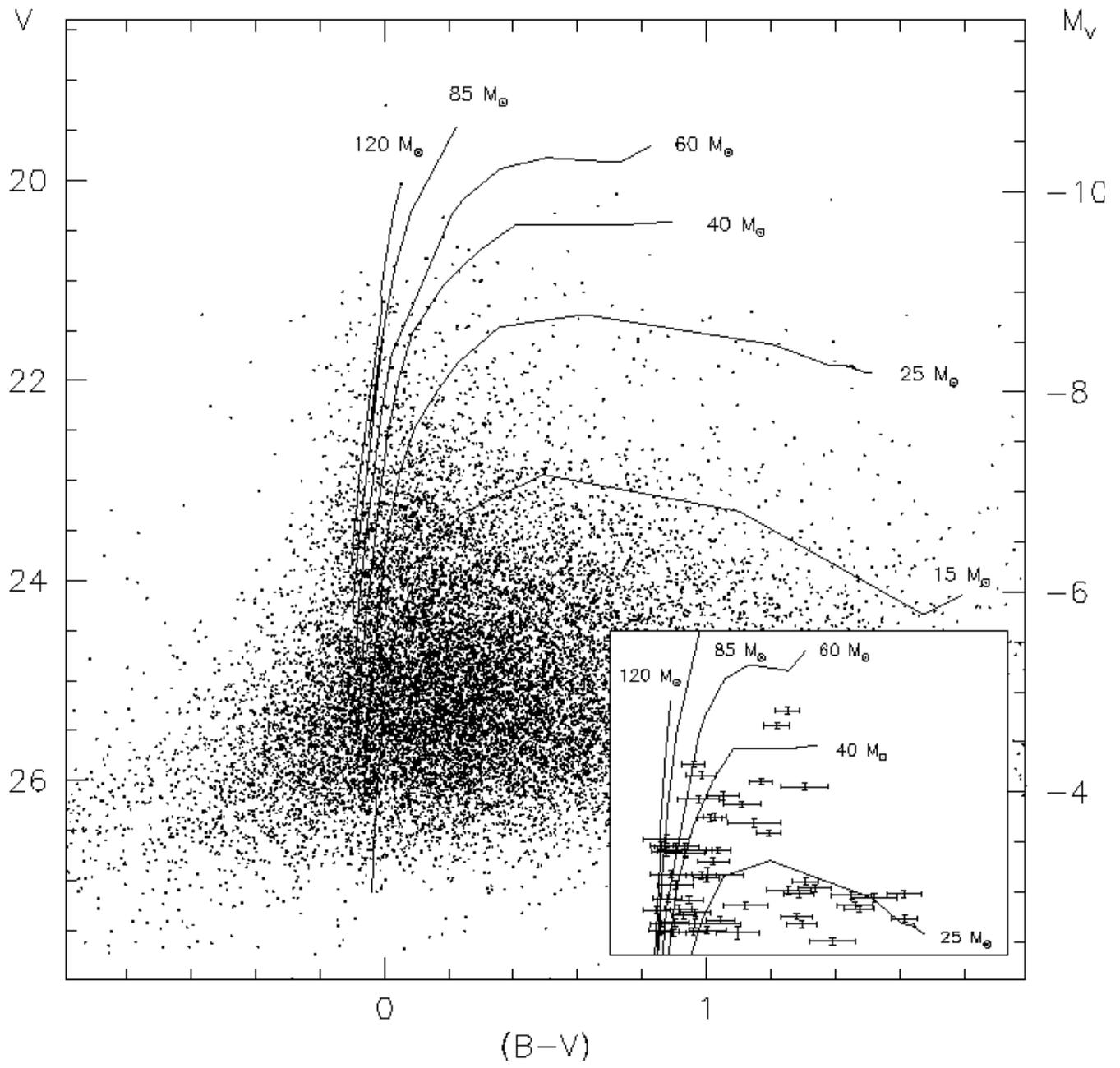
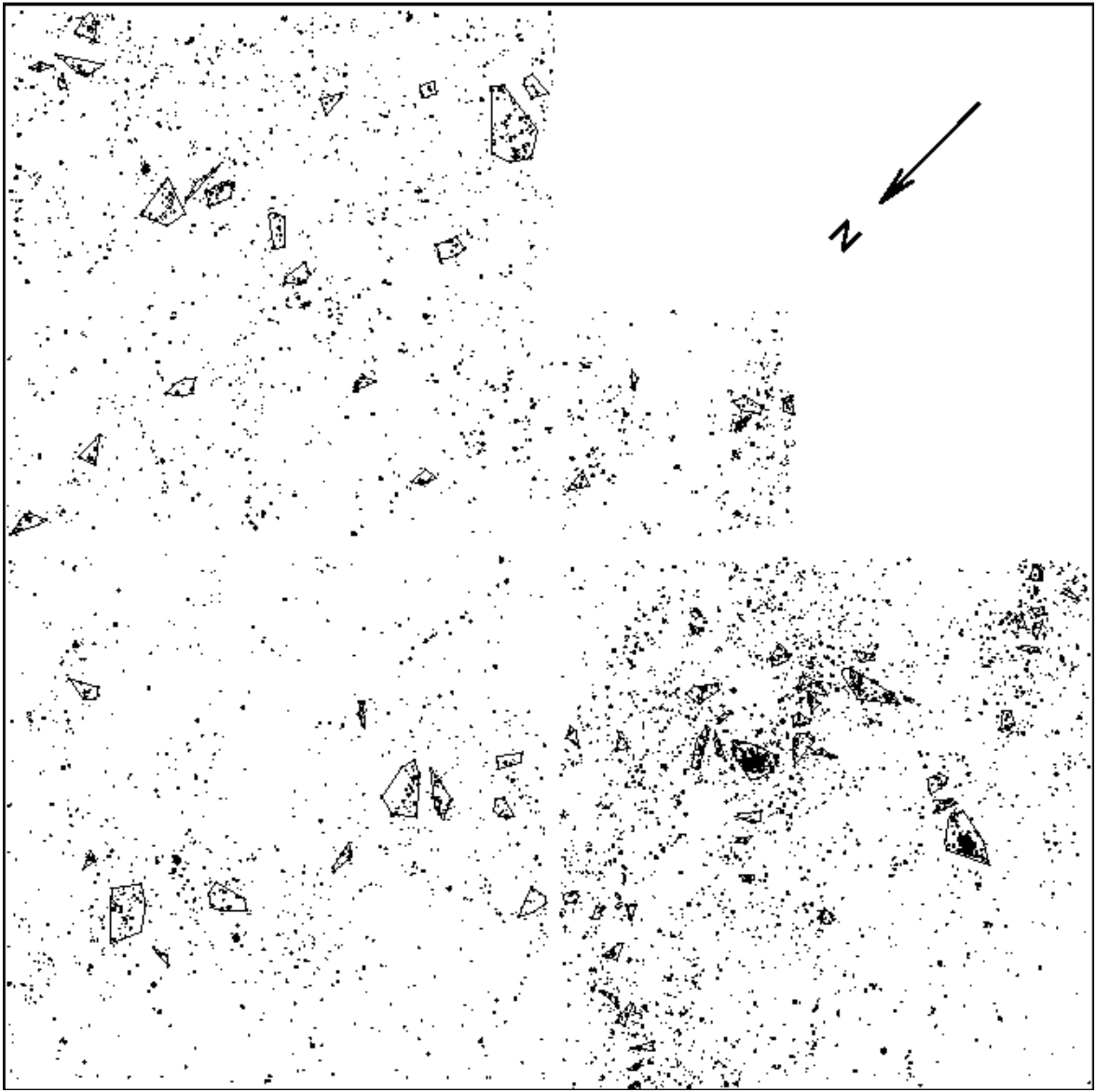
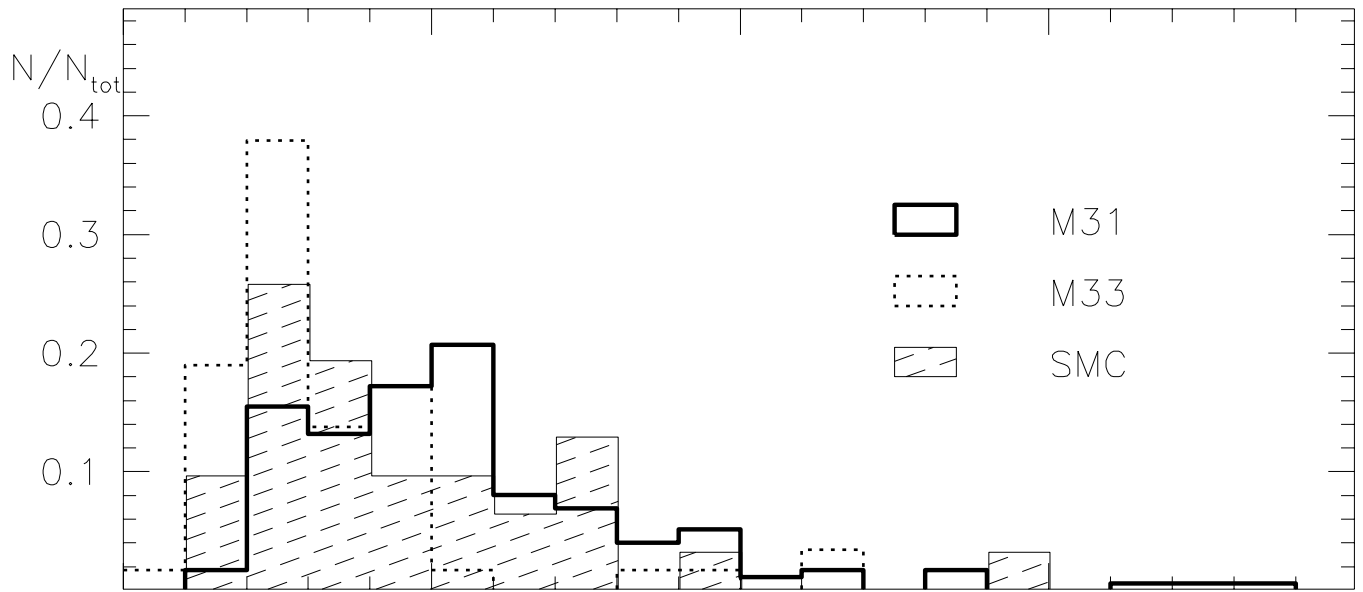
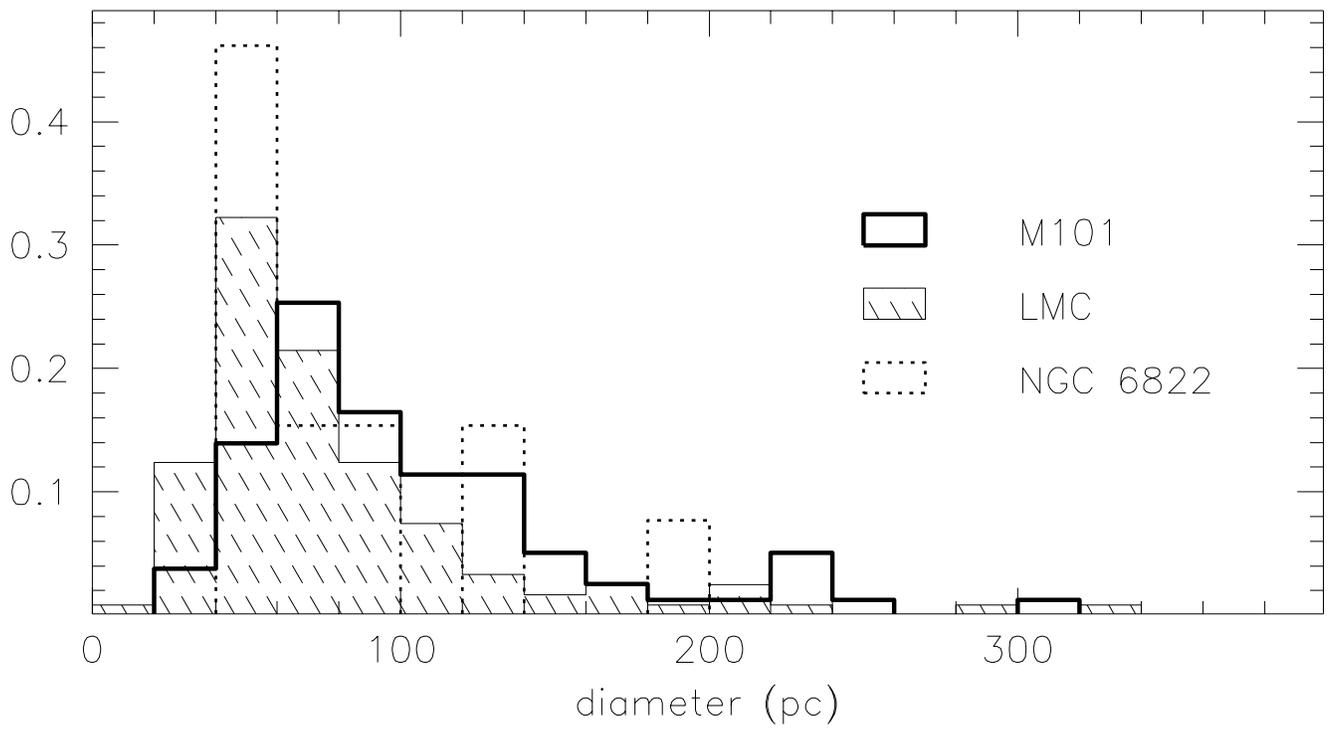
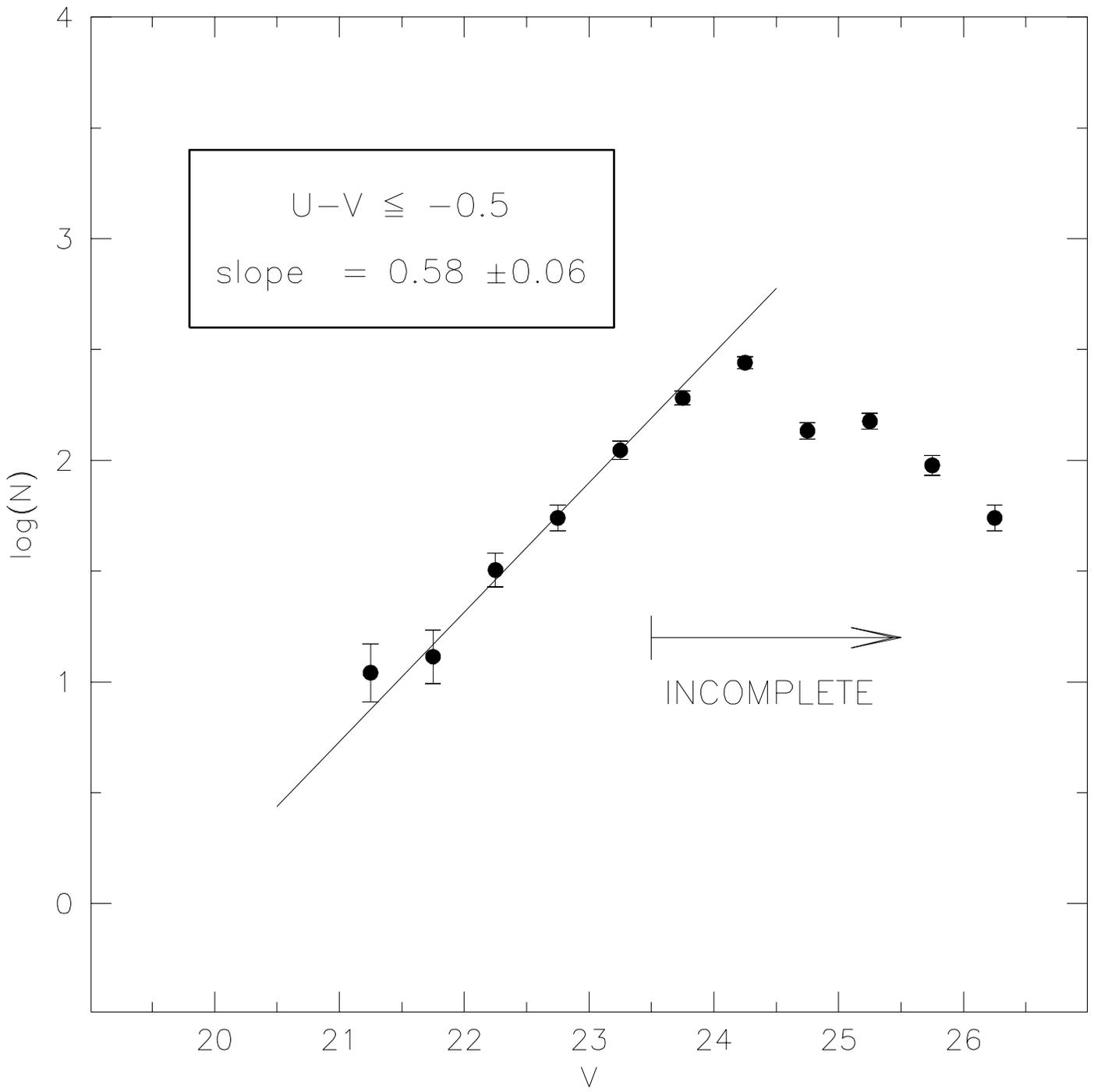
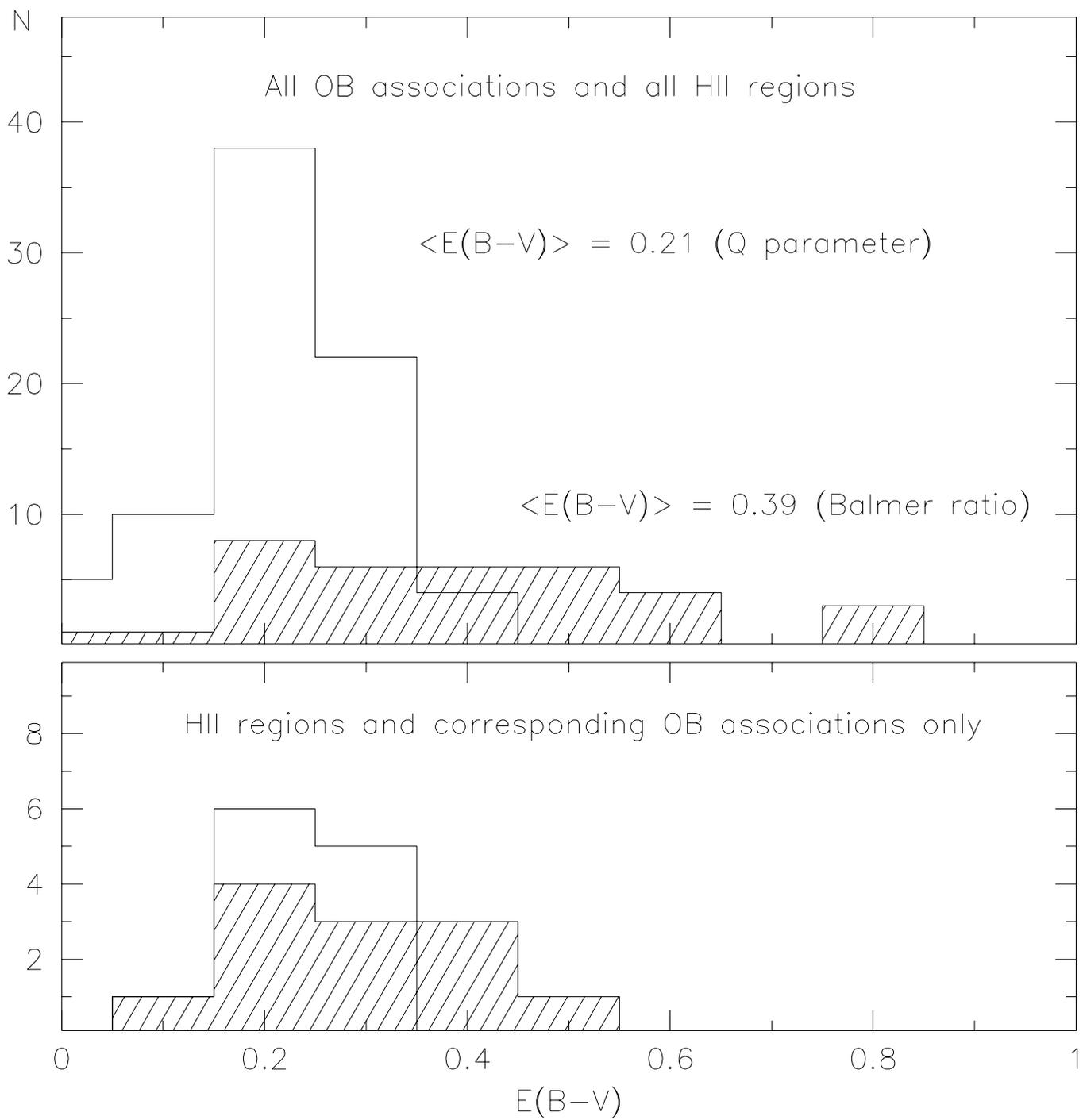
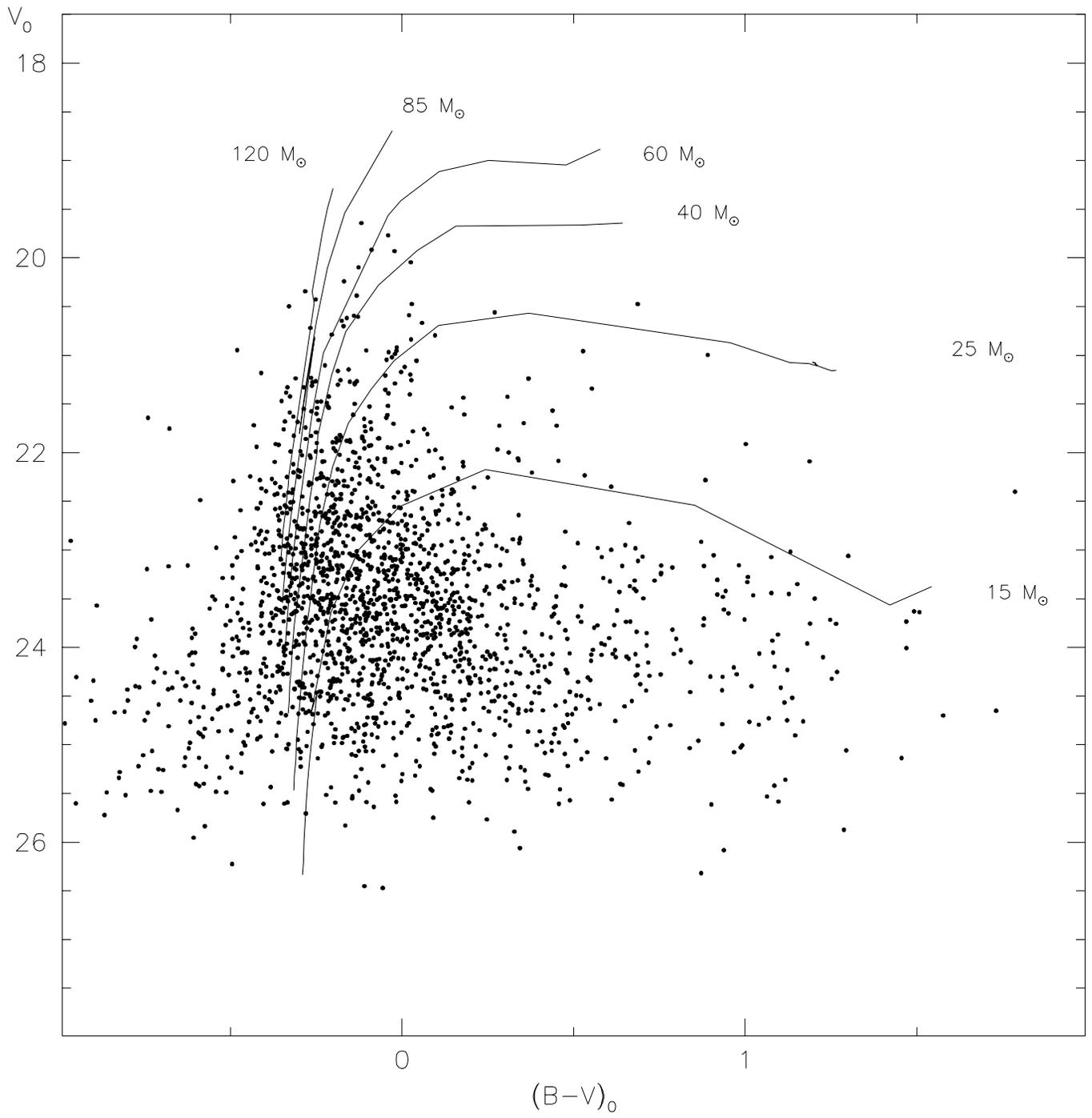
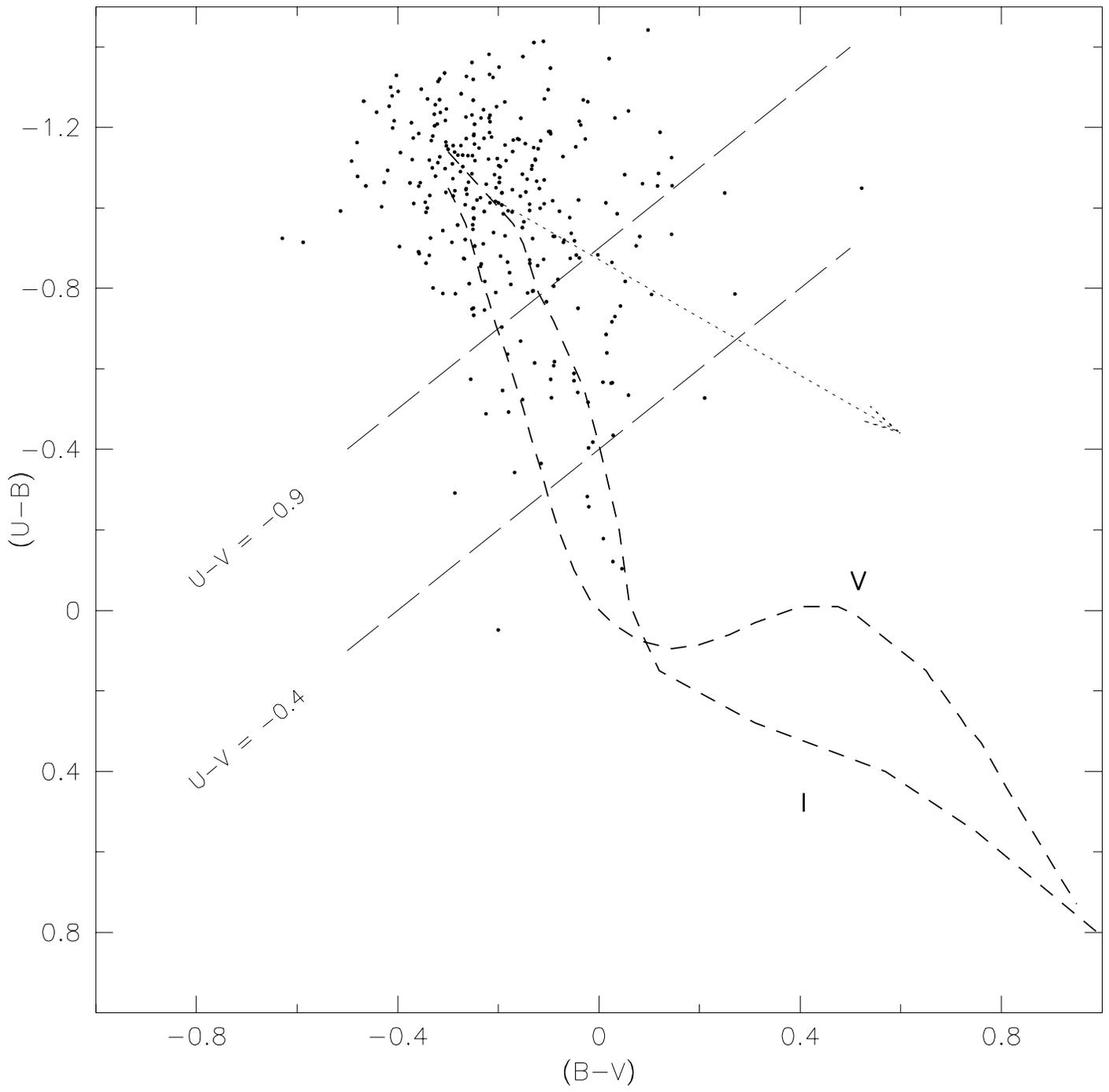
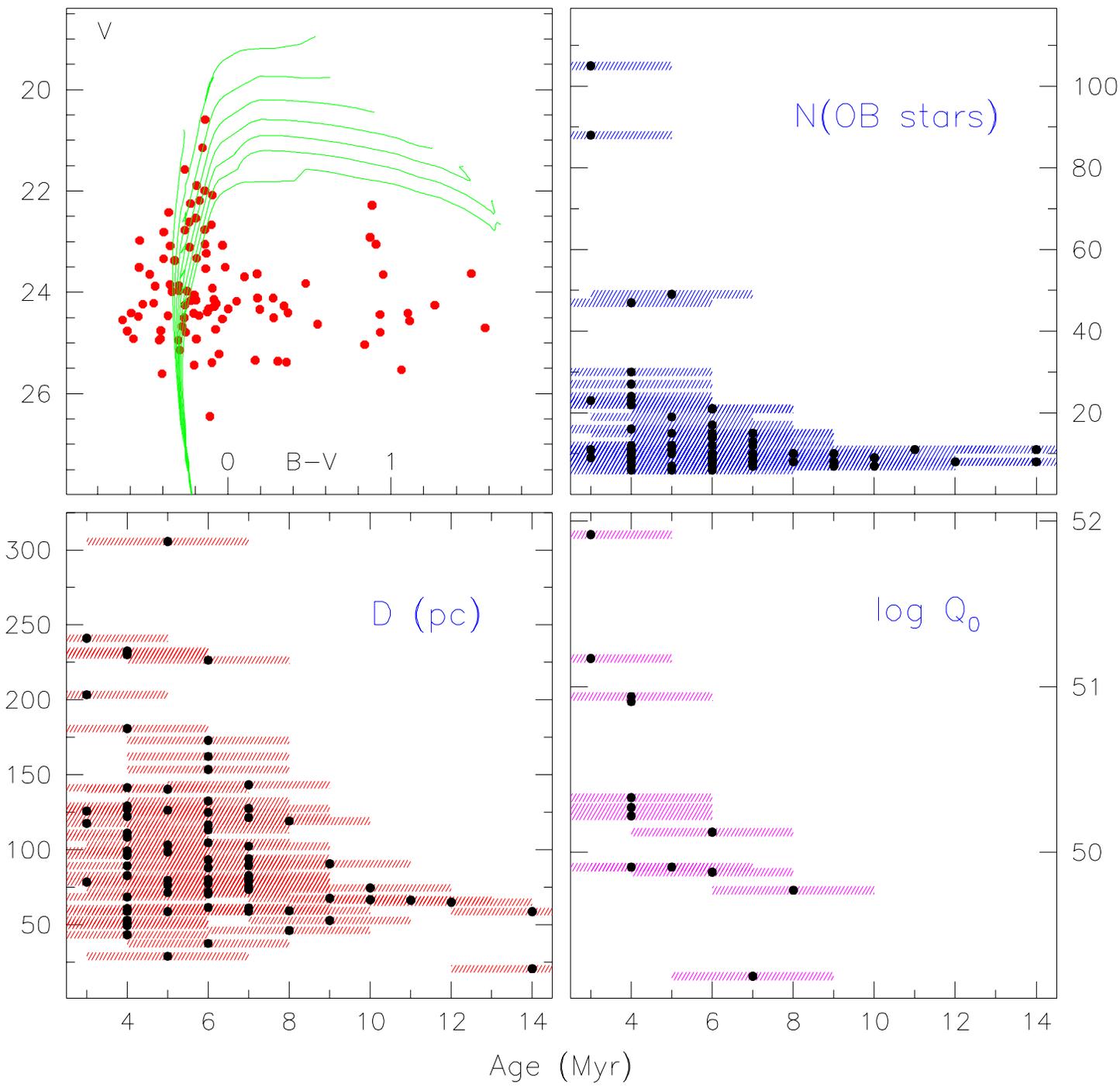
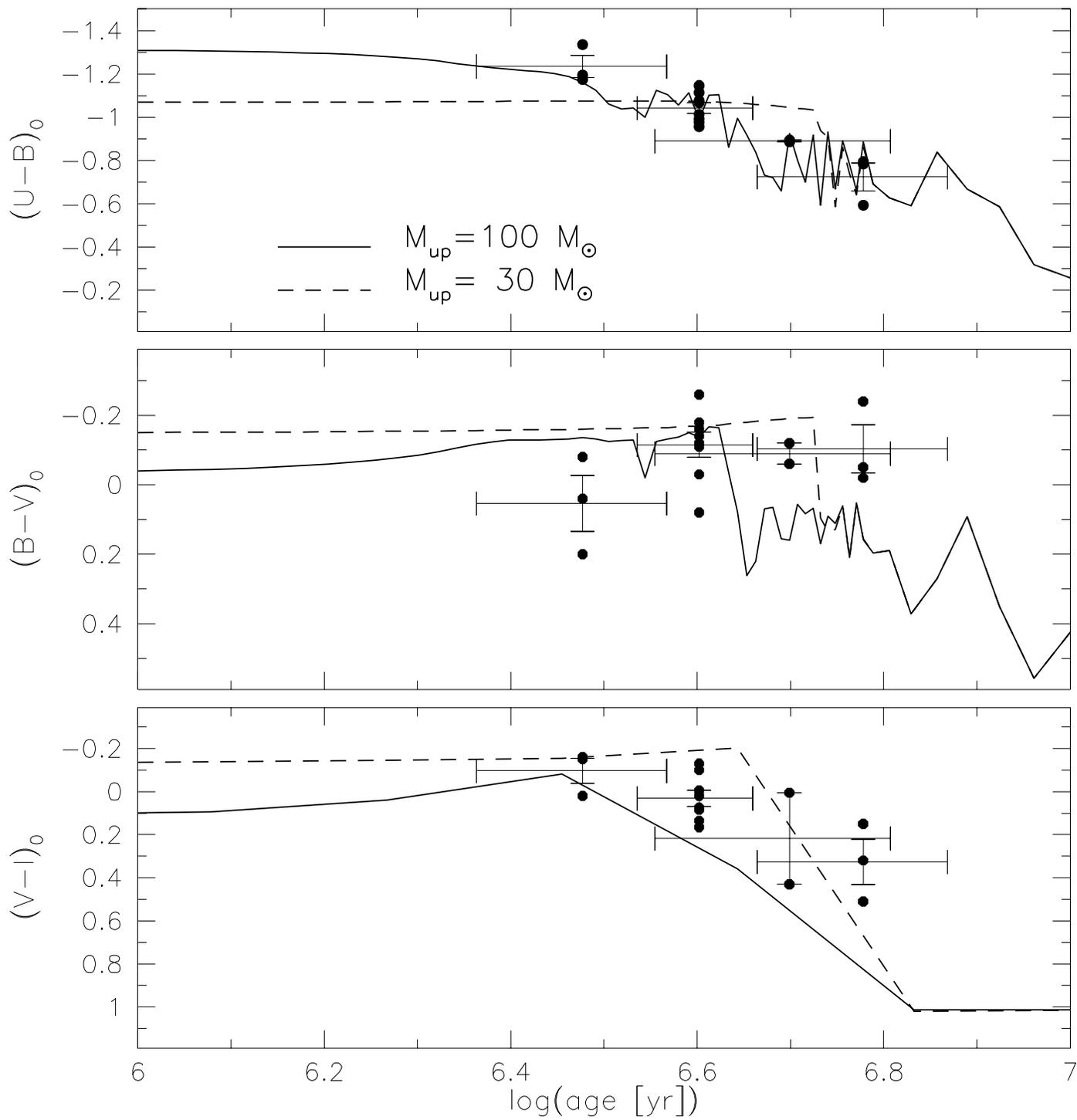
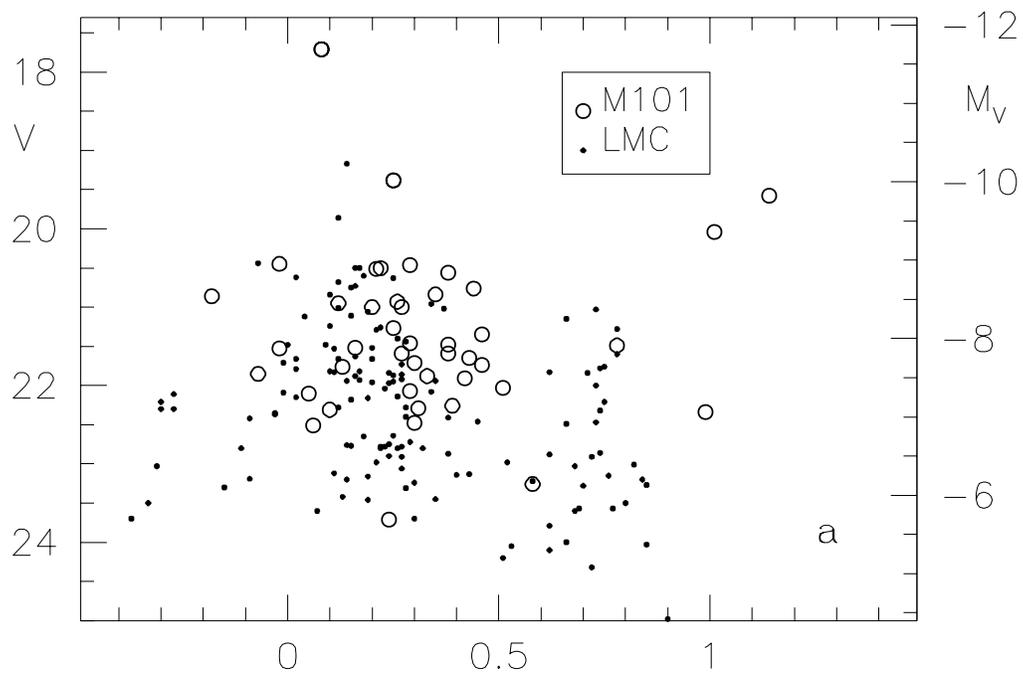
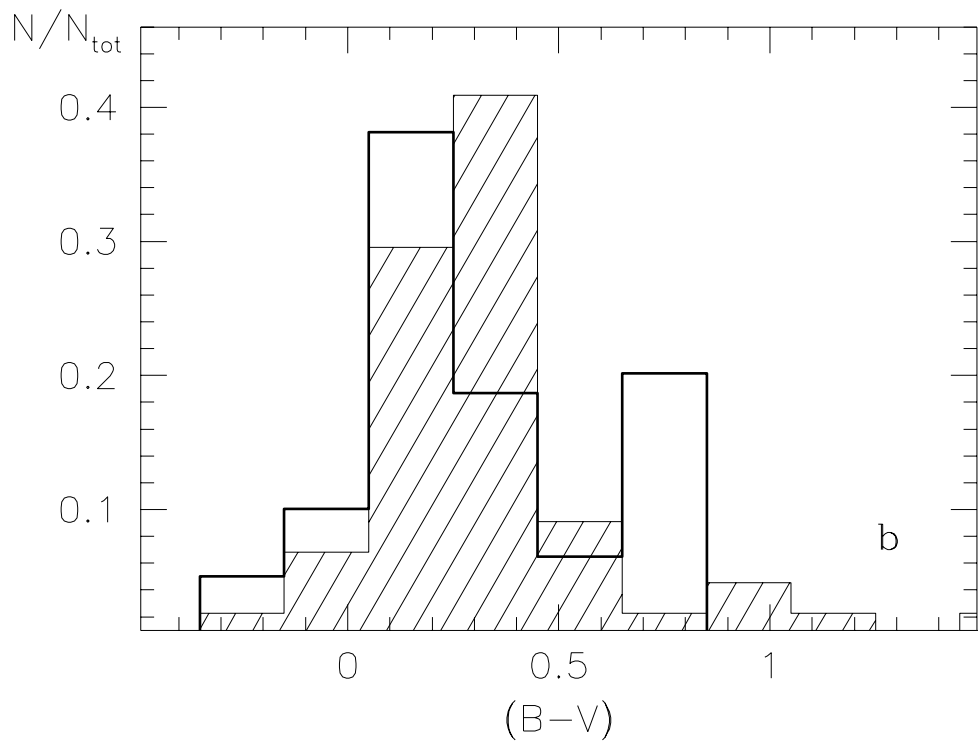
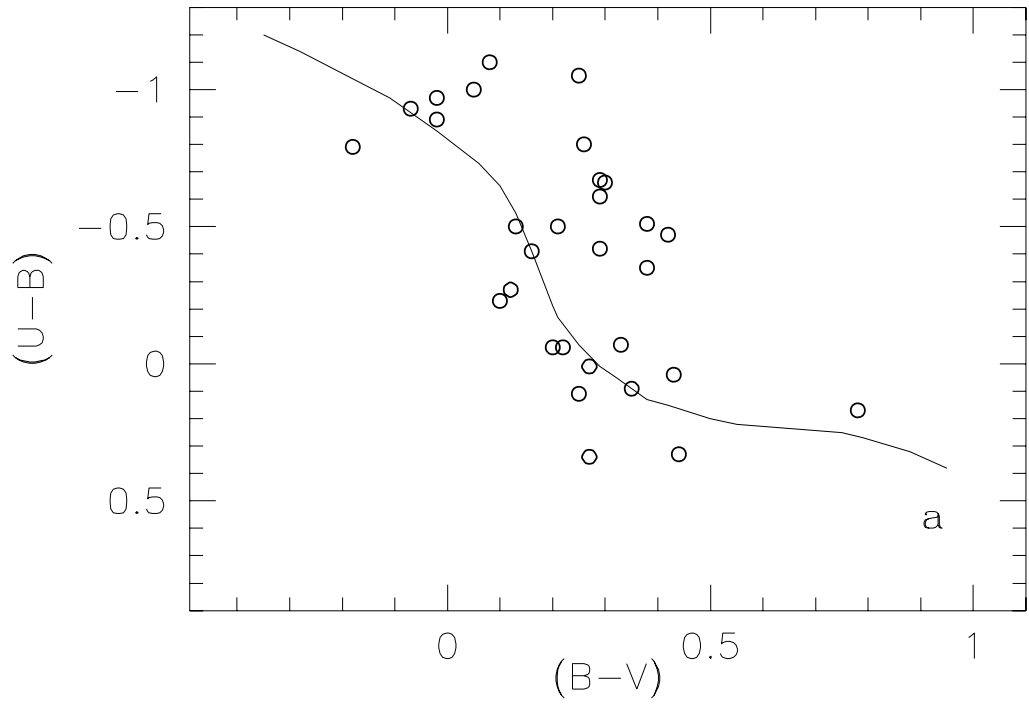
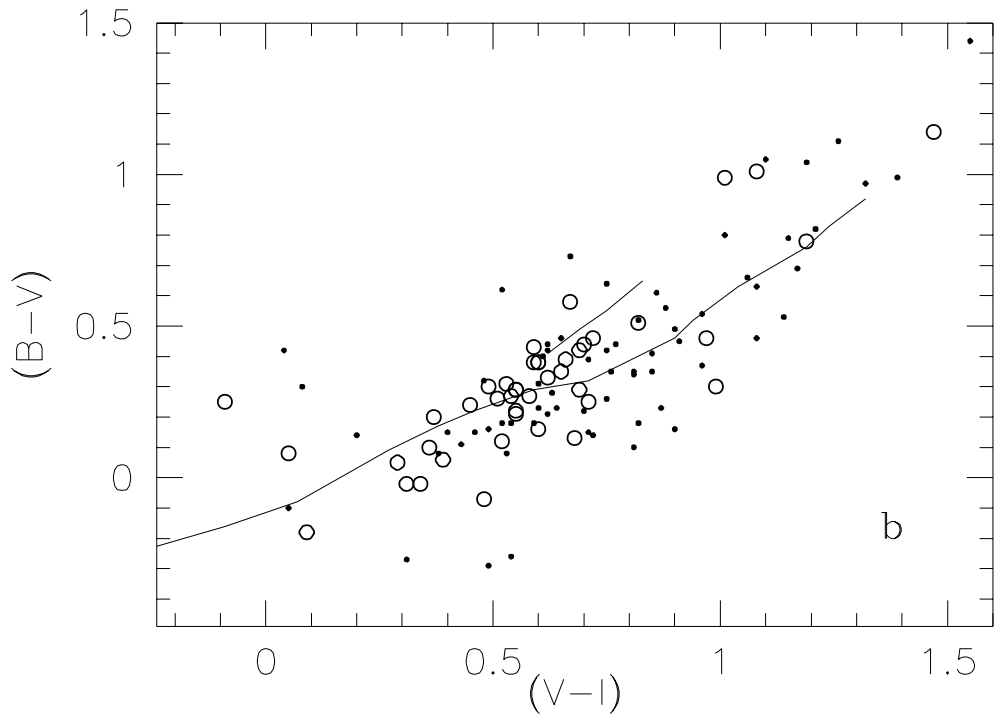
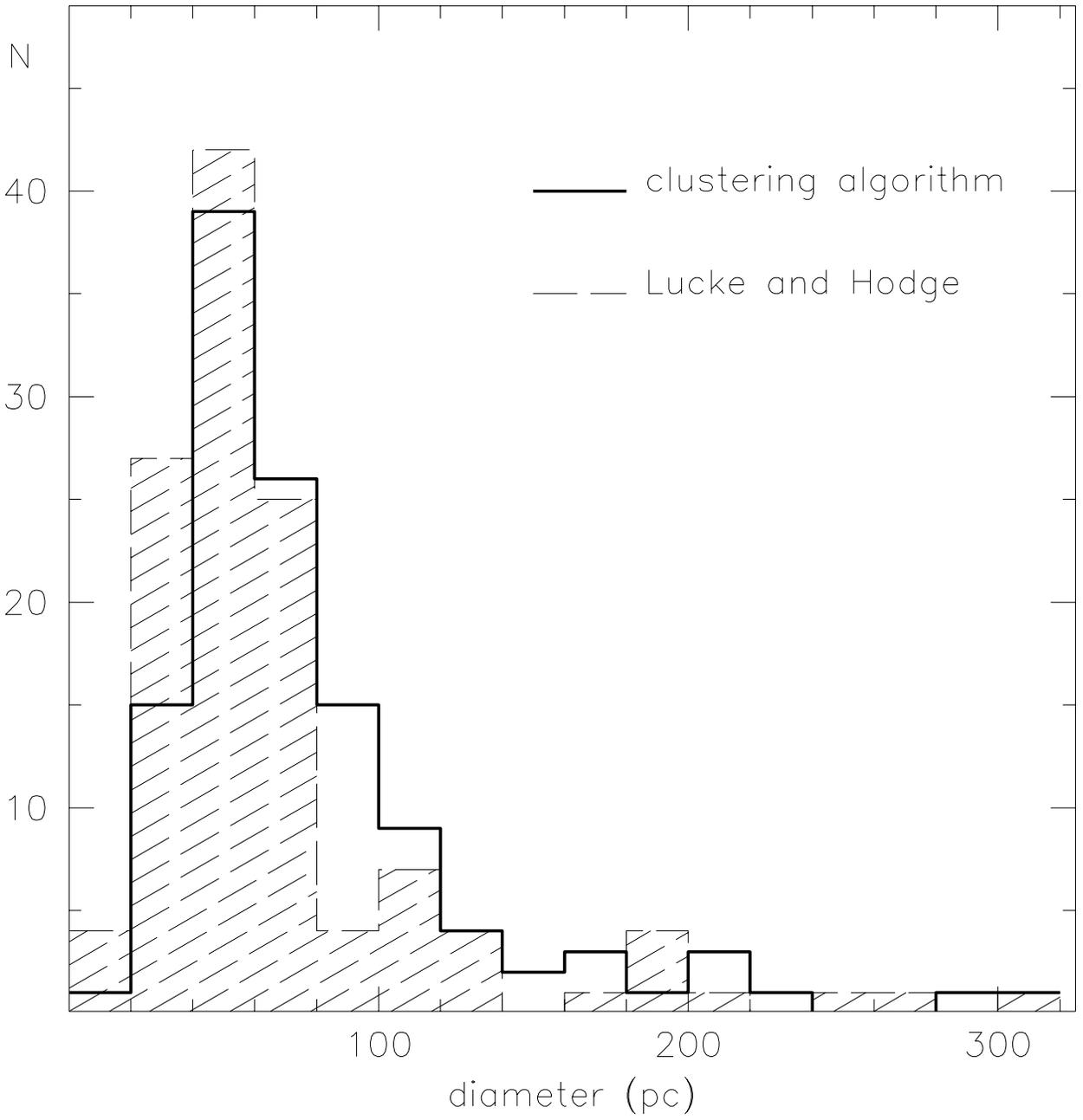
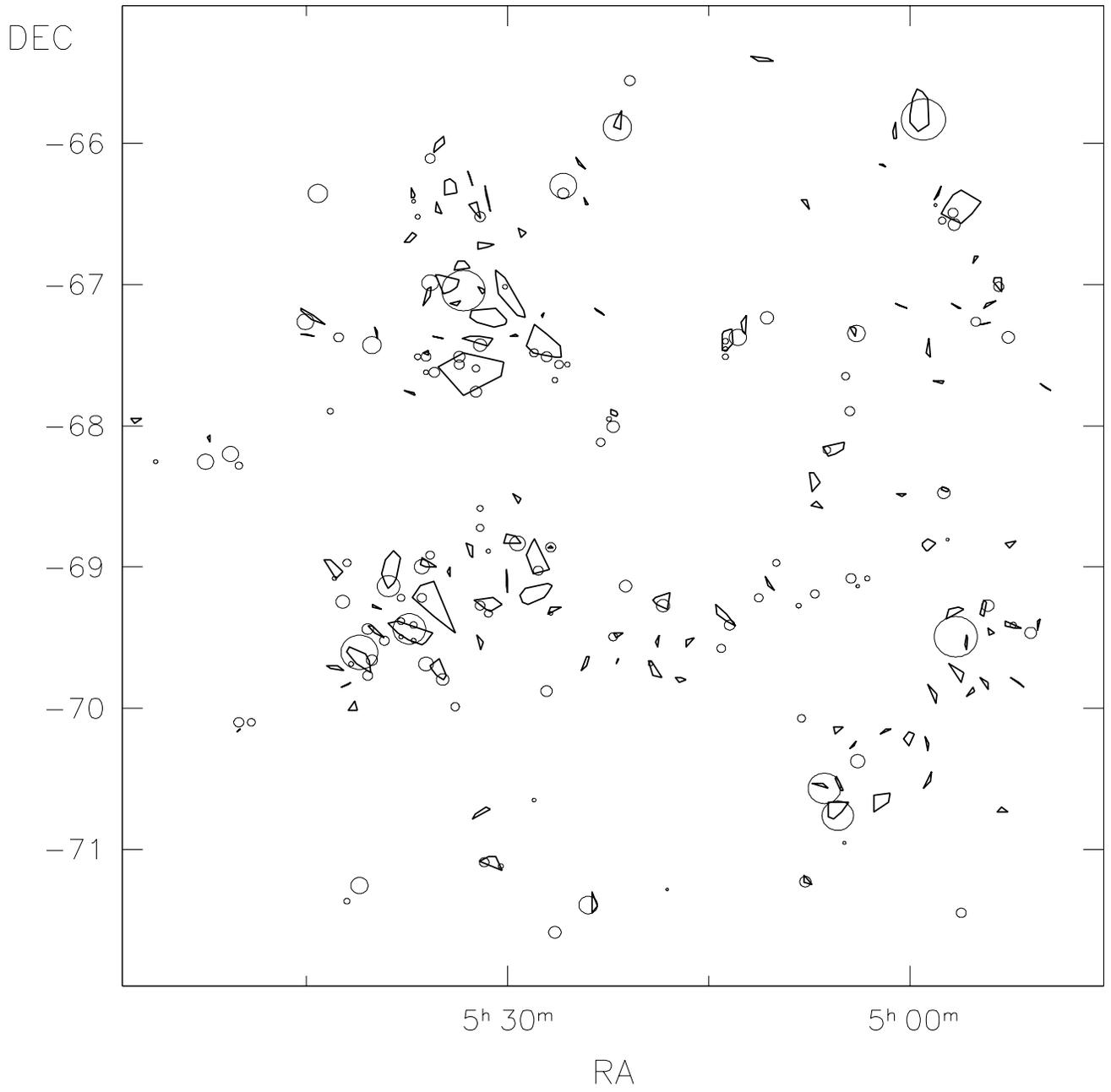
\end{document}